\documentclass{article}

\usepackage{graphicx}
\usepackage{amssymb}
\usepackage{amsmath}
\usepackage{amsthm}
\usepackage{mathrsfs}
\usepackage{url}
\usepackage{color}
\usepackage{longtable}
\usepackage{rotating}
\usepackage{authblk}
\usepackage{multirow}

\definecolor{red}{rgb}{1,0,0}

\begin{document}

\title{Quasi-Centralized Limit Order Books}

\author{Martin D. Gould\footnote{Corresponding author. Email: m.gould@imperial.ac.uk.} \footnote{MDG also completed part of this work while at the University of Oxford.} $^{\ddag}$, Mason A. Porter$^{\S \P \ast\ast}$, and Sam D. Howison$^{\S \|}$}

\affil{$\ddag$ CFM--Imperial Institute of Quantitative Finance, Imperial College, London SW7 2AZ\\$\S$ Oxford Centre for Industrial and Applied Mathematics, Mathematical Institute, University of Oxford, Oxford OX2 6GG\\$\P$ CABDyN Complexity Centre, University of Oxford, Oxford OX1 1HP\\ $\ast\ast$ Department of Mathematics, University of California, Los Angeles, California 90095 \\ $\|$ Oxford--Man Institute of Quantitative Finance, University of Oxford, Oxford, OX2 6ED}

\date{ }

\maketitle

\begin{abstract}A quasi-centralized limit order book (QCLOB) is a limit order book (LOB) in which financial institutions can only access the trading opportunities offered by counterparties with whom they possess sufficient bilateral credit. In this paper, we perform an empirical analysis of a recent, high-quality data set from a large electronic trading platform that utilizes QCLOBs to facilitate trade. We argue that the quote-relative framework often used to study other LOBs is not a sensible reference frame for QCLOBs, so we instead introduce an alternative, trade-relative framework, which we use to study the statistical properties of order flow and LOB state in our data. We also uncover an empirical universality: although the distributions that describe order flow and LOB state vary considerably across days, a simple, linear rescaling causes them to collapse onto a single curve. Motivated by this finding, we propose a semi-parametric model of order flow and LOB state for a single trading day. Our model provides similar performance to that of parametric curve-fitting techniques but is simpler to compute and faster to implement.\end{abstract}


\section{Introduction}

More than half of the world's financial markets use electronic limit order books (LOBs) to facilitate trade \cite{Rosu:2009dynamic}. In contrast to quote-driven systems, in which prices are set by designated market makers, trades in an LOB occur via a continuous double-auction mechanism, in which institutions submit orders that state their desire to buy or sell a specified quantity of an asset at a specified price. Active orders reside in a queue until they are either cancelled by their owner or executed against an order of opposite type. Upon execution, the owners of the relevant orders trade the agreed quantity of the asset at the agreed price.

During the past 20 years, a large body of empirical and theoretical work has addressed a specific type of LOB in which all institutions are able to trade with all others (see \cite{Gould:2013limit} for a review). We call this market organization a \emph{centralized LOB}. Although several large platforms -- including the London Stock Exchange (LSE) Electronic Trading Service \cite{LSEWebsite}, Nasdaq \cite{NASDAQLOBWebsite}, and the Euronext Universal Trading Platform \cite{EuronextWebsite} -- employ centralized LOBs, many other platforms use alternative LOB configurations. In contrast to the wealth of publications on centralized LOBs, discussion of alternative LOB configurations is limited to a handful of technical descriptions of matching mechanisms on specific platforms \cite{Barker:2007global,Gallardo:2009execution,Rime:2003new,Sarno:2001microstructure}. Given their widespread use, detailed study of alternative LOB configurations is an important task.

A prominent example of an alternative LOB configuration is an LOB in which financial institutions can only access the trading opportunities offered by counterparties with whom they possess sufficient bilateral credit. We call this market organization a \emph{quasi-centralized limit order book (QCLOB)} because different institutions have access to different subsets of a centralized liquidity pool. QCLOBs are used by several major multi-institution trading platforms in the foreign exchange (FX) spot market, including Reuters \cite{ReutersWebsite}, EBS \cite{EBS:2011}, and Hotspot FX \cite{HotspotWebsite}, which together facilitate a mean turnover in excess of $0.6$ trillion US dollars (USD) each day \cite{BIS:2010triennial}.

Despite this enormous volume of trade, a lack of adequate data has hindered investigation of many important questions regarding QCLOBs. Do the statistical properties of QCLOBs differ from those of centralized LOBs? Do arbitrage opportunities arise? How do institutions assess market state when deciding how to act? In this paper, we present an empirical study of a recent, high-quality data set from Hotspot FX that enables us to address these issues.

In comparison to the statistics that are widely reported in empirical studies of centralized LOBs (see \cite{Gould:2013limit}), we observe much lower levels of order flow at the prevailing quotes and a much higher ratio of active liquidity to market order flow. We also identify periods during which the global bid--ask spread is negative. Due to the extremely high levels of market activity on Hotspot FX, we are able to perform both cross-sectional (i.e., between different currency pairs) and longitudinal (i.e., across different time periods) comparisons of our findings. We find several longitudinal differences in market activity, and we thus argue that using long-run statistical averages to formulate short-run forecasts may produce misleading results. We also uncover a striking empirical universality: applying a simple, linear rescaling to the distributions that describe order flow and market state causes the data to collapse onto a single curve. Motivated by this finding, we propose a semi-parametric model of these distributions that gives similar performance to parametric curve-fitting techniques but is simpler to compute and faster to implement.

Our findings are important for several reasons. First, they provide a detailed overview of recent trading activity on a large electronic trading platform. Second, they illustrate similarities and differences between market activity on different trading days. Third, they highlight how several properties of QCLOBs differ from those of centralized LOBs. Fourth, they motivate a semi-parametric model for the distributions that describe order flow and market state in a QCLOB. Together, our results help to illuminate the delicate interplay between order flow, liquidity, and price formation for a widely used but hitherto unexplored market organization.

The paper proceeds as follows. In Section \ref{sec:lobs}, we present several definitions that we use throughout the paper, provide a detailed description of centralized LOBs and QCLOBs, and highlight the important differences between these mechanisms. In Section \ref{sec:hotspotdata}, we describe the data that forms the basis for our empirical study and discuss the Hotspot FX platform. In Section \ref{sec:methodology}, we describe the methodology that we use for our empirical study. We present our main results in Section \ref{sec:results} and discuss our findings in Section \ref{sec:sfxdiscussion}. We conclude in Section \ref{sec:conclusions}. In Appendix \ref{app:fittingt}, we describe our method of performing parametric fits to daily data. In Appendix \ref{app:curvecollapse}, we describe how we quantify the strength of curve collapse when rescaling each day's data in our semi-parametric model.

\section{Centralized and Quasi-Centralized Limit Order Books}\label{sec:lobs}

Let $\Theta=\left\{\theta_1,\theta_2,\ldots\right\}$ denote the set of institutions that trade a given asset on a given platform. In an LOB, these institutions interact by submitting orders. An \emph{order} $x=(p_x,\omega_x,t_x)$ submitted at time $t_x$ with price $p_x$ and size $\omega_x>0$ (respectively, $\omega_x<0$) is a commitment by its owner to sell (respectively, buy) up to $\left|\omega_x\right|$ units of the asset at a price no less than (respectively, no greater than) $p_x$.

Whenever an institution submits a buy (respectively, sell) order $x$, an LOB's trade-matching algorithm checks whether it is possible for $x$ to \emph{match to} an active sell (respectively, buy) order $y$ such that $p_y \leq p_x$ (respectively, $p_y \geq p_x$). If so, the matching occurs immediately and the owners of the relevant orders agree to trade the specified amount at the specified price. If $\left|\omega_x\right|>\left|\omega_y\right|$, any residue of $x$ is then considered for matching to other active sell (respectively, buy) orders until either $x$ becomes fully matched or there are no further active sell (respectively, buy) orders eligible for matching to $x$. Any portion of $x$ that does not match becomes active at the price $p_x$, and it remains active until it either matches to an incoming sell (respectively, buy) order or is \emph{cancelled}.

Orders that match completely upon arrival are called \emph{market orders}. Orders that do not match upon arrival --- instead becoming active in the LOB --- are called \emph{limit orders}.\footnote{Some orders match partially upon arrival. Such orders can be construed as partly a market order and partly a limit order.} Some platforms allow other order types -- such as fill-or-kill, stop-loss, or peg orders \cite{HotspotGUIUserGuide} -- but it is always possible to decompose the resulting order flow into limit and/or market orders. Therefore, we study LOBs in terms of these simple building blocks.

The \emph{global}\footnote{We use the term ``global'' to highlight the differences between these definitions and the local definitions in Section \ref{subsec:qclobs}.} \emph{LOB} $\mathcal{L}(t)$ is the set of all active orders for a given asset on a given platform at time $t$. The \emph{global bid price} $b(t)$ is the highest price among active buy orders in $\mathcal{L}(t)$. The \emph{global ask price} $a(t)$ is the lowest price among active sell orders in $\mathcal{L}(t)$. The \emph{global bid--ask spread} is $s(t)=a(t)-b(t)$. The \emph{global mid price} is $m(t)=\left[b(t)+a(t)\right]/2$.

\subsection{Centralized LOBs}\label{subsec:clobs}

In a \emph{centralized LOB}, all institutions can trade with all others. Whenever an institution $\theta_i$ submits a buy (respectively, sell) market order, the order matches to the highest-priority active sell (respectively, buy) order that is owned by another institution $\theta_j$, irrespective of the identities of $\theta_i$ and $\theta_j$. Therefore, all institutions in a centralized LOB face the same trading opportunities. A sell order with $p_x > b(t)$ or a buy order with $p_x < a(t)$ is always a limit order, a sell order with arbitrarily small $p_x$ or a buy order with arbitrarily large $p_x$ is always a market order, and a sell order with $p_x \leq b(t)$ or a buy order with $p_x \geq a(t)$ at least partially matches immediately upon arrival. For a detailed discussion of centralized LOBs, see \cite{Gould:2013limit}.

\subsection{Quasi-Centralized LOBs}\label{subsec:qclobs}

In a QCLOB, each institution can specify the maximum level of counterparty credit exposure that it is willing to extend to each other institution trading on the platform.\footnote{In the FX spot market, trades agreed on day $d$ are settled on day $d+2$. Therefore, each trade by an institution in this market entails exposure to the counterparty during the period between trade agreement and trade settlement. Mitigation of the resulting counterparty risk is one reason for the use of CCLs.} Specifically, each institution $\theta_i$ in a QCLOB notifies the exchange of its \emph{counterparty credit limit (CCL) $c_{(i;j)}\geq 0$} for each other institution $\theta_j$. Assigning a CCL to a given counterparty does not require posting collateral; instead, it simply involves notifying the exchange of the relevant value $c_{(i,j)}$. Institution $\theta_i$ cannot access any trading opportunities offered by another institution $\theta_j$ that would make $\theta_i$'s total exposure to $\theta_j$ exceed $c_{(i,j)}$ or that would make $\theta_j$'s total exposure to $\theta_i$ exceed $c_{(j,i)}$. Hence the maximal amount that $\theta_i$ and $\theta_j$ can trade is $\min\left(c_{(i,j)},c_{(j,i)}\right)$. We call this quantity the \emph{bilateral CCL between $\theta_i$ and $\theta_j$.} The bilateral CCLs determine the subset of trading opportunities available to each institution. This subset changes over time according to the relevant institutions' trading activity.

Institution $\theta_i$ can ensure that it never trades with $\theta_j$ by setting $c_{(i,j)}= 0$, because arranging any trade with $\theta_j$ would result in a non-zero exposure and would thereby violate this CCL. Institution $\theta_i$ can also assign an unlimited amount of credit to $\theta_j$ by setting $c_{(i,j)}= \infty$. Irrespective of the CCL set by $\theta_i$, it still remains open to $\theta_j$ to further restrict the bilateral exposure by choosing $c_{(j,i)}$ appropriately. In particular, the choice $c_{(j;i)}= 0$ indicates unwillingness to trade at all.

\begin{figure}[!htbp]
\centering
\includegraphics[width=0.3\textwidth]{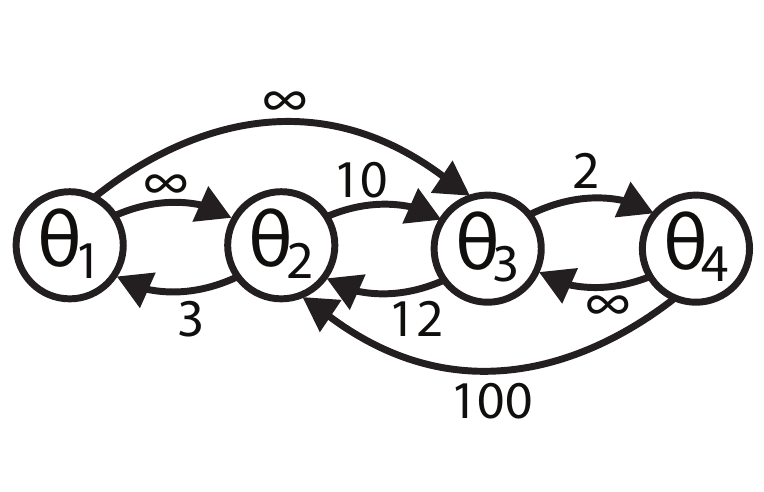} \\
\includegraphics[width=0.3\textwidth]{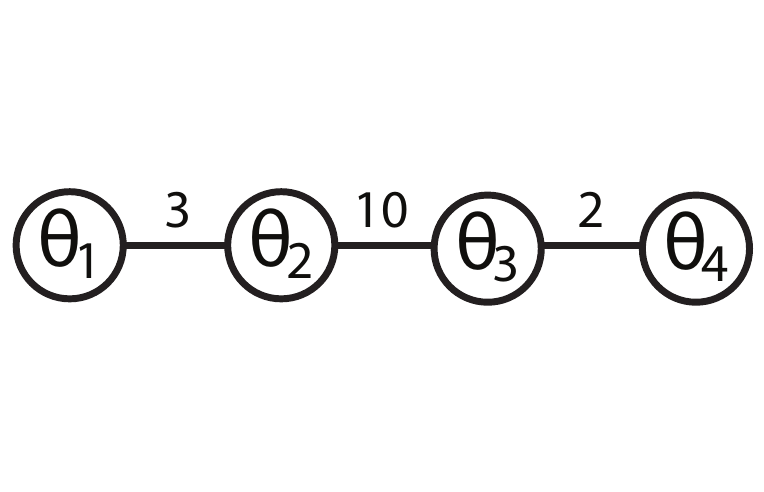}
\caption{Two weighted network representations of the CCLs in a QCLOB; see the main text for details. (Top) Directed network with edge weights equal to the corresponding CCLs. (Bottom) Undirected network with edge weights equal to the corresponding bilateral CCLs. In both networks, edges with zero weight are omitted.}
\label{fig:networks}
\end{figure}

In Figure \ref{fig:networks}, we show two possible network representations of the CCLs in a QCLOB populated by institutions $\Theta=\left\{\theta_1,\theta_2,\theta_3,\theta_4\right\}$ with CCLs\begin{equation}\label{eq:example_ccls}\begin{aligned}
    c_{(1,2)}=\infty,\qquad &c_{(1,3)}=\infty,\\
    c_{(2,1)}=3,\qquad &c_{(2,3)}=10,\\
		c_{(3,2)}=12,\qquad &c_{(3,4)}=2,\\
    c_{(4,2)}=100,\qquad &c_{(4,3)}=\infty,\end{aligned}\end{equation}and with all other CCLs equal to 0. In both representations, nodes corresponds to institutions and edge weights to CCLs. The first representation is a directed network in which the weight of the edge from node $i$ to node $j$ is equal to the CCL $c_{(i,j)}$. The second representation is an undirected network in which the weight of the edge between nodes $i$ and $j$ is equal to the bilateral CCL between institutions $i$ and $j$ (i.e., $\min\left(c_{(i,j)},c_{(j,i)}\right)$). In this example, the CCL structure is akin to a core of two creditworthy institutions ($\theta_2$ and $\theta_3$), which can trade freely with each other, and two peripheral, less creditworthy, institutions ($\theta_1$ and $\theta_4$), each of which can only trade with one core partner.

\begin{figure}[!htbp]
\centering
\includegraphics[width=0.5\textwidth]{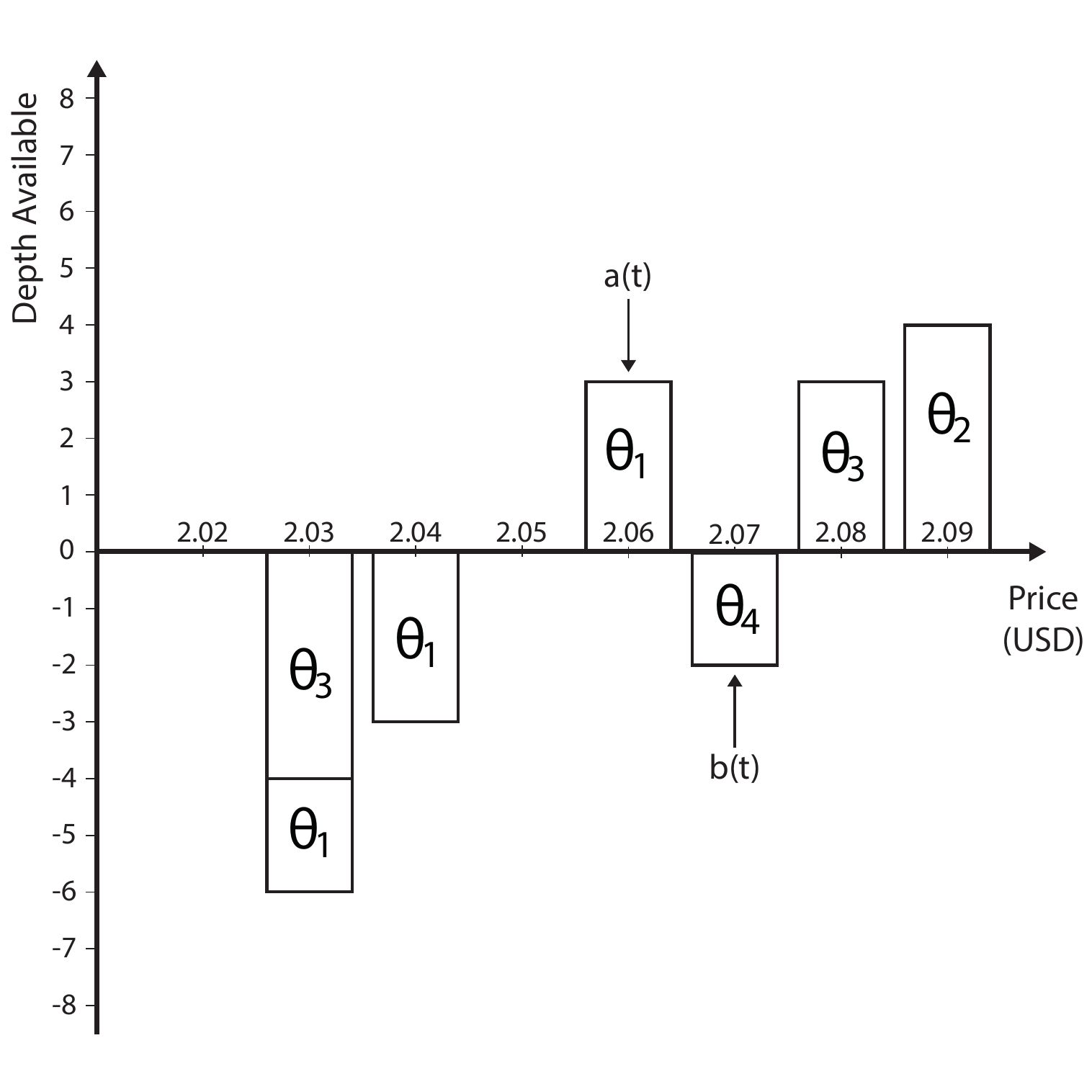} \\
\includegraphics[width=0.4\textwidth]{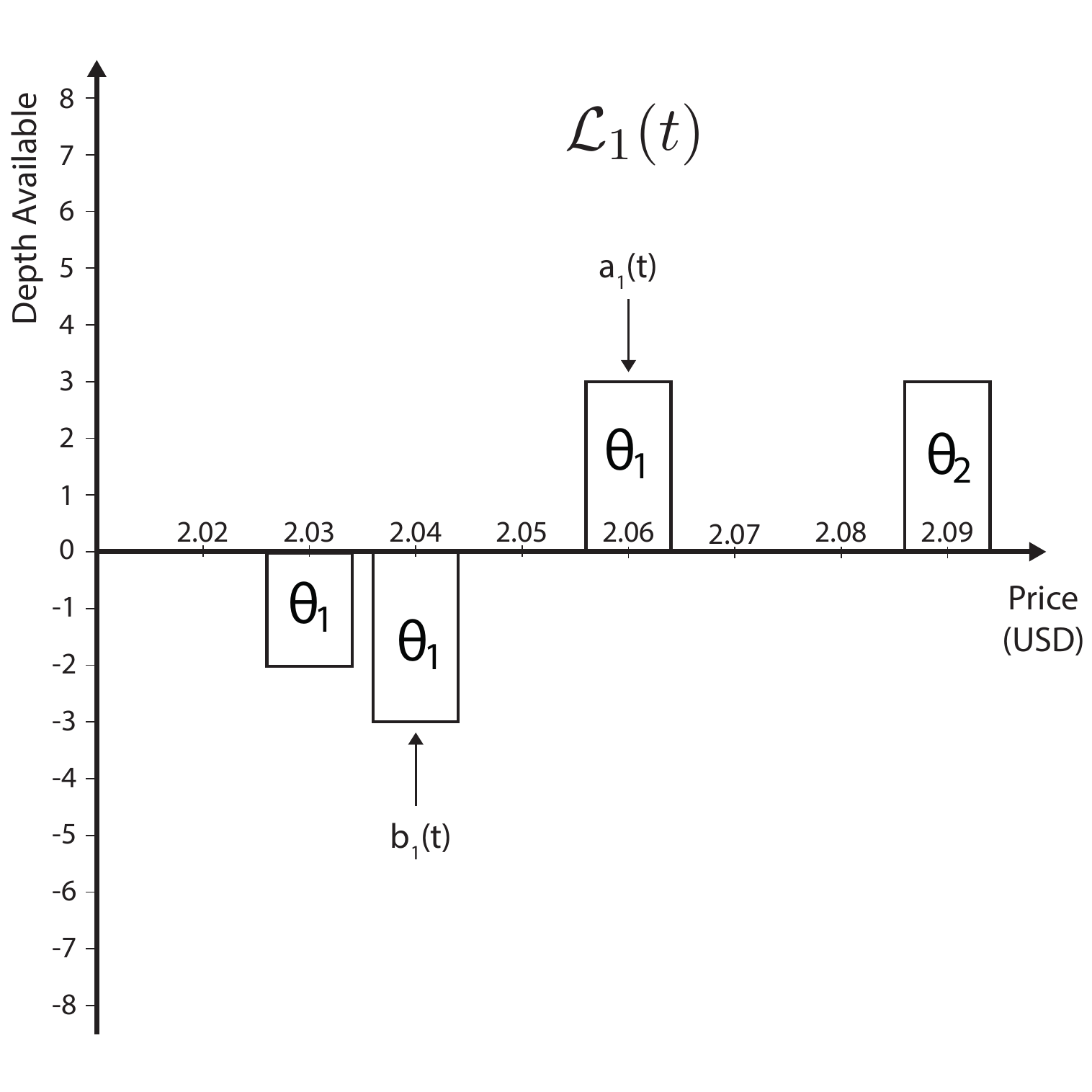}
\includegraphics[width=0.4\textwidth]{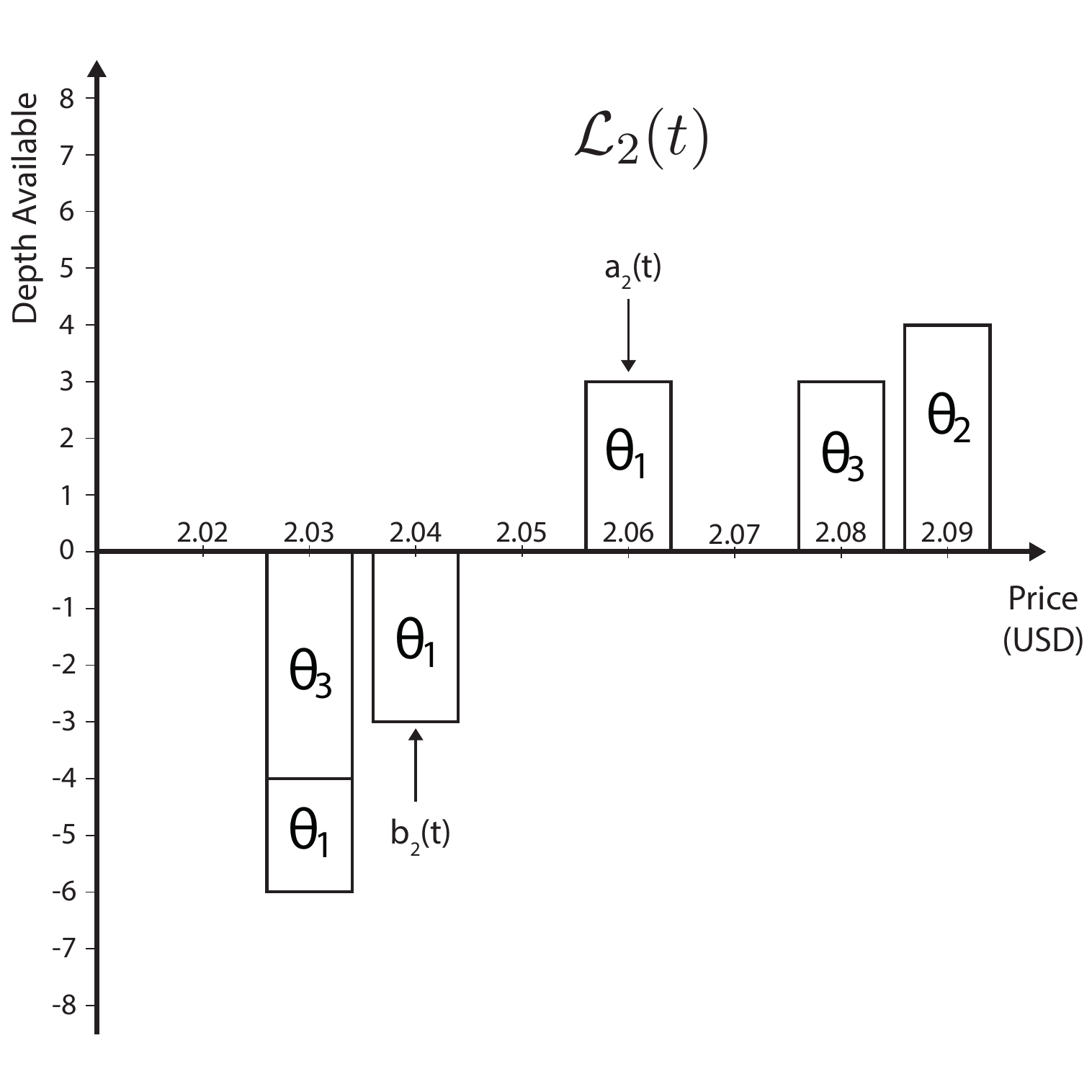} \\
\includegraphics[width=0.4\textwidth]{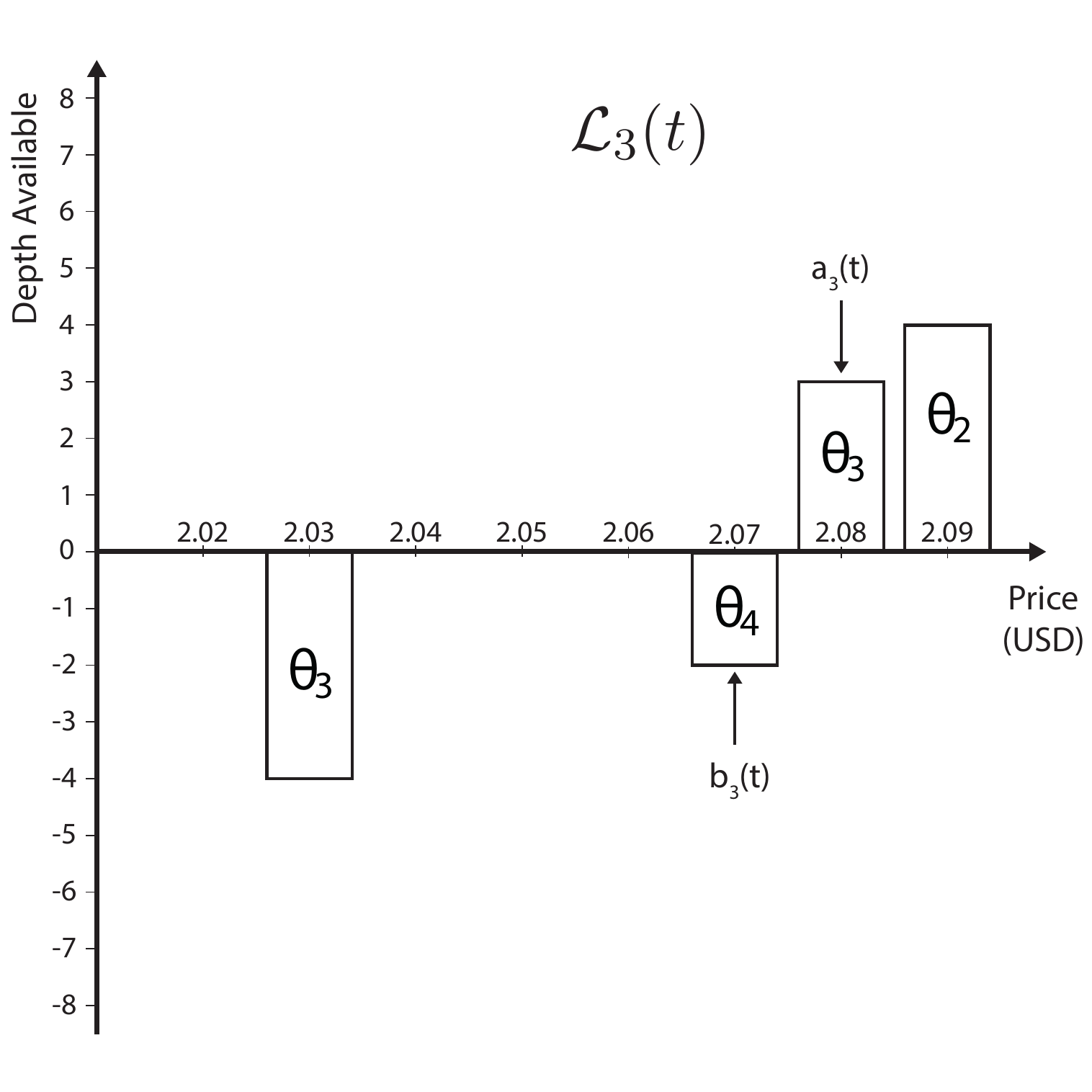}
\includegraphics[width=0.4\textwidth]{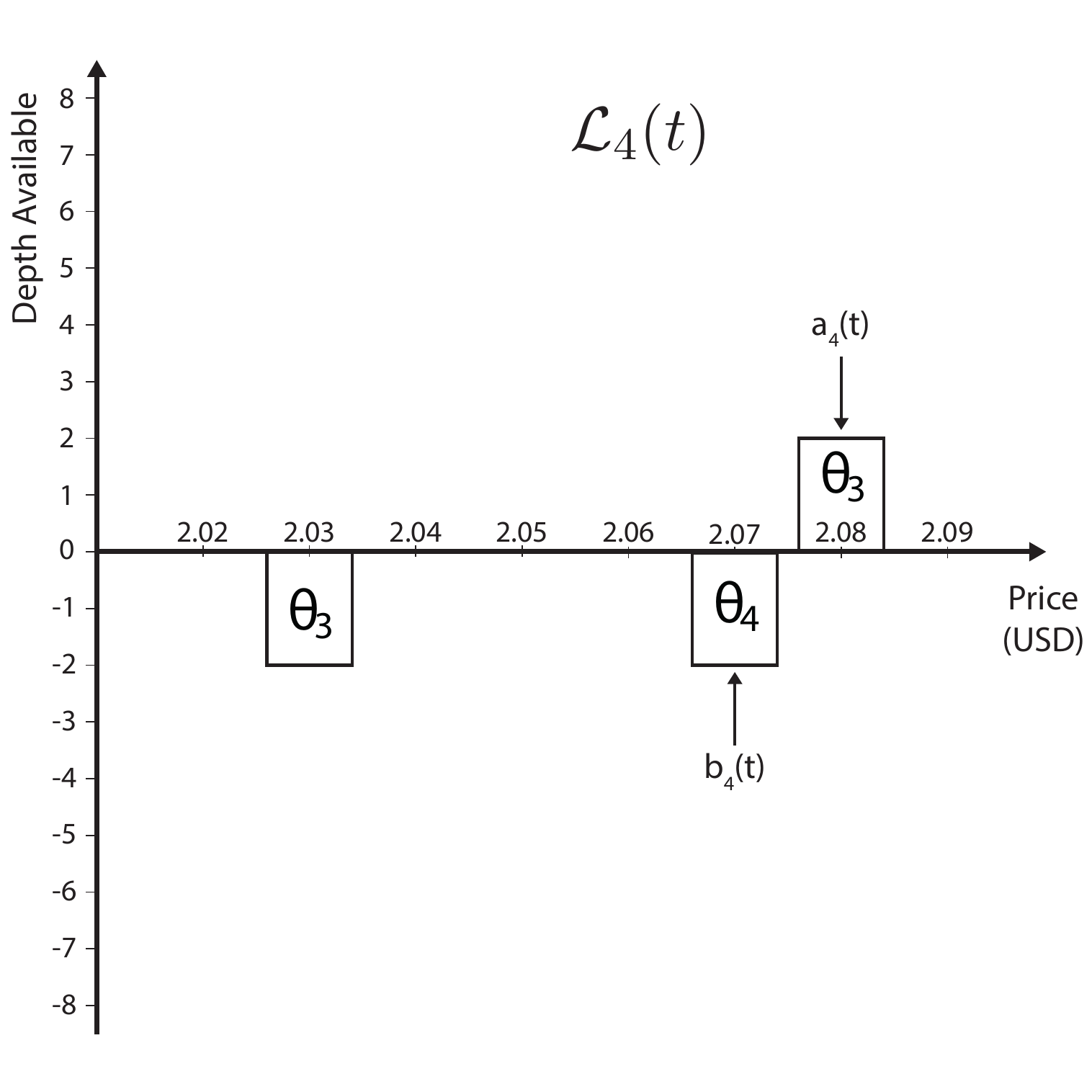}
\caption{Schematic of (top) a global LOB $\mathcal{L}(t)$ and (top left) $\theta_1$'s local LOB $\mathcal{L}_1(t)$, (top right) $\theta_2$'s local LOB $\mathcal{L}_2(t)$, (bottom left) $\theta_3$'s local LOB $\mathcal{L}_3(t)$, and (bottom right) $\theta_4$'s local LOB $\mathcal{L}_3(t)$ for a QCLOB with the CCLs described in Figure \ref{fig:networks}. To illustrate the role of CCLs, we label each order in the figure according to its owner. However, trading platforms do not disseminate this information.}
\label{fig:unfilteredfiltered}
\end{figure}

Institutions trading on a QCLOB platform cannot in general see the state of the global LOB $\mathcal{L}(t)$. Instead, each institution $\theta_i$ sees only the active orders that correspond to trading opportunities that it can access (i.e., do not violate any of its bilateral CCLs) at time $t$.\footnote{Some QCLOB platforms (such as Reuters and EBS) offer institutions the ability to access an additional data feed that provides snapshots of the global LOB $\mathcal{L}(t)$ at regular time intervals in exchange for a fee.} This filtering of $\mathcal{L}(t)$ yields local versions of several key concepts (see Figure \ref{fig:unfilteredfiltered}). Institution $\theta_i$'s \emph{local LOB} $\mathcal{L}_i(t)$ is the subset of active orders in $\mathcal{L}(t)$ that $\theta_i$ can access. More precisely, for each $j\neq i$, the volume of each separate limit order placed by $\theta_j$ is reduced (if necessary) in $\mathcal{L}_i(t)$ so that its size does not exceed the bilateral CCL between $\theta_i$ and $\theta_j$.

Institution $\theta_i$'s \emph{local bid price} $b_i(t)$ is the highest stated price among active buy orders in $\mathcal{L}_i(t)$. Institution $\theta_i$'s \emph{local ask price} $a_i(t)$ is the lowest stated price among active sell orders in $\mathcal{L}_i(t)$. Institution $\theta_i$'s \emph{local bid--ask spread} is $s_i(t)=a_i(t)-b_i(t)$. Institution $\theta_i$'s \emph{local mid price} is $m_i(t)=\left[b_i(t)+a_i(t)\right]/2$.

When an institution $\theta_i$ submits a buy (respectively, sell) market order, the order matches to the highest-priority active sell (respectively, buy) order in $\mathcal{L}_i(t)$. Importantly, there may be higher-priority active sell (respectively, buy) orders in the global LOB $\mathcal{L}(t)$ owned by another institution $\theta_j$ with whom $\theta_i$ has insufficient bilateral credit to perform the trade, but such orders are not considered for matching to $\theta_i$'s market order because they do not appear in $\theta_i$'s local LOB $\mathcal{L}_i(t)$.

A noteworthy difference between a QCLOB and a centralized LOB follows from the partial nature of each institution's local LOB. In a QCLOB, the global spread $s(t)$ (which is observable in our data) can be negative even though the local spreads $s_i(t)$ (which are not observable in our data) are positive. In Section~\ref{sec:results}, we report that negative global spreads occur reasonably frequently, but do not persist for long.

In addition to viewing their local LOB $\mathcal{L}_i(t)$, each institution in a QCLOB can access a \emph{trade-data stream} that lists the price, time, and direction (buy/sell) of each trade that occurs. All institutions can see all entries in the trade-data stream in real time, irrespective of their bilateral CCLs with the institutions involved in a given trade. Therefore, although institutions in a QCLOB do not have access to information regarding which trading opportunities are available to other institutions, they do have access to a detailed historical record of previous trades.

In Figure~\ref{fig:unfilteredfiltered}, we illustrate an example of a QCLOB's global and local LOBs. The figure shows a simple global LOB and the corresponding local LOBs for the four institutions shown in Figure~\ref{fig:networks}. In the figure, we label each order according to its owner, although this information is not visible to traders. In this example, the global spread is negative, but all local spreads are positive. Observe that in $\mathcal{L}_1(t)$, the order owned by $\theta_2$ is truncated to size $3$, because this is the value of the bilateral CCL between $\theta_1$ and $\theta_2$. Similarly, in $\mathcal{L}_4(t)$, the orders owned by $\theta_3$ are truncated to size $2$, because this is the value of the bilateral CCL between $\theta_3$ and $\theta_4$.

\subsection{Coordinate Frames}\label{subsec:coordinateframes}

Because a financial institution's activity is driven by its trading needs, its individual actions can appear extremely erratic. However, many empirical studies of centralized LOBs (see, e.g., \cite{Biais:1995empirical,Bouchaud:2002statistical,Chakraborti:2011empirical,Challet:2001analyzing,Cont:2010stochastic,Gu:2008empiricalregularities,Gu:2008empiricalshape,Mike:2008empirical,Potters:2003more,Zovko:2002power}) have noted that when measured in a suitable coordinate frame that aggregates order flows from many different institutions, robust statistical properties can emerge from the ensemble.

Most studies of centralized LOBs perform such aggregation in a coordinate frame that we call \emph{quote-relative coordinates}, in which prices are measured relative to the global bid price $b(t)$ or the global ask price $a(t)$. Specifically, the \emph{quote-relative price of an order $x$ at time $t$} is\begin{equation}\label{eq:quoterelative}\phi(p_x,t) := \left\{ \begin{array}{ll}
b(t)-p_x, & \text{if }x\text{ is a buy order,} \\
p_x-a(t), & \text{if }x\text{ is a sell order.}
\end{array}\right.\end{equation}The difference in signs between the definitions for buy and sell orders ensures that all active orders have a non-negative quote-relative price at all times.

The use of quote-relative coordinates in centralized LOBs is motivated by the notion that institutions monitor $b(t)$ and $a(t)$ when deciding how to act. There are many reasons why this is the case. For example, $b(t)$ and $a(t)$ define the boundary conditions that dictate whether an incoming order is a limit order or a market order, they are observable to all institutions in real time, and they are common to all institutions. Therefore, they constitute suitable reference points for aggregating order flows across different institutions.

In a QCLOB, by contrast, the boundary conditions between limit order and market order placement for a given institution $\theta_i$ are determined by $\theta_i$'s local bid price $b_i(t)$ and local ask price $a_i(t)$, rather than the global values $b(t)$ and $a(t)$. Moreover, institutions cannot see the state of the global LOB $\mathcal{L}(t)$, so they do not know the values of $b(t)$ and $a(t)$. Therefore, quote-relative coordinates are not a natural framework for studying QCLOBs. This provides strong motivation to explore alternative avenues.

Given complete information regarding each institution's local LOB $\mathcal{L}_i(t)$, one possible approach would be to measure each institution's order flow relative to its local quotes $b_i(t)$ and $a_i(t)$ and to aggregate the corresponding relative prices across institutions. However, this approach would require calculating each institution's local LOB $\mathcal{L}_i(t)$, which is not possible using the Hotspot FX data (see Section \ref{subsec:data}). Another alternative is to measure all institutions' order flow relative to a benchmark price that is common to all institutions and visible to all institutions in real time. Recall from Section \ref{subsec:qclobs} that QCLOBs disseminate a trade-data stream that lists the prices of all previous trades. This trade-data stream thereby facilitates the use of an alternative coordinate frame, which we call \emph{trade-relative coordinates}, in which prices are measured relative to those of the most recent trades. Let $B(t)$ and $A(t)$ denote, respectively, the price of the most recent seller-initiated and buyer-initiated trades (across all institutions) that occur at or before time $t$. The \emph{trade-relative price of an order $x$ at time $t$} is then given by
\begin{equation}\label{eq:traderelative}\Phi(p_x,t) := \left\{ \begin{array}{ll}
B(t)-p_x, & \text{if }x\text{ is a buy order,} \\
p_x-A(t), & \text{if }x\text{ is a sell order.}
\end{array}\right.\end{equation}

In contrast to quote-relative prices, all institutions in a QCLOB can calculate trade-relative prices in real time. Moreover, we can calculate trade-relative prices directly from our Hotspot FX data (see Section \ref{subsec:data}). Therefore, trade-relative coordinates are a useful alternative to quote-relative coordinates in a QCLOB.

To highlight their similarities and differences, we perform our calculations throughout the paper in both quote-relative and trade-relative coordinates. We find that using quote-relative coordinates produces relatively weak statistical signals with high variance, but that using trade-relative coordinates helps to uncover stable and robust statistical regularities.

\section{Hotspot FX}\label{sec:hotspotdata}

\subsection{The Hotspot FX Platform}\label{subsec:hfx}

We have been granted access to a recent, high-quality data set from Hotspot FX \cite{HotspotWebsite,HotspotGUIUserGuide}, which is one of the largest multi-institution trading platforms in the FX spot market. The data describes all limit order arrivals, cancellations, and trades during May--June 2010. According to the 2010 Triennial Central Bank Survey \cite{BIS:2010triennial}, the mean daily turnover of the global FX market around this time was approximately $4.0$ trillion USD. Approximately $37\%$ of this volume was due to spot trades, of which approximately $40\%$ was conducted electronically. In total, the mean daily volume traded on all multi-institution electronic trading platforms was approximately $0.6$ trillion USD \cite{BIS:2010triennial}. The mean daily volume traded on Hotspot FX during the same period was approximately $21.5$ billion USD \cite{HotspotVolumes}. Therefore, trade on Hotspot FX accounted for approximately $4\%$ of all volume traded electronically in the FX spot market during this period.

Hotspot FX offers trade for more than 60 different currency pairs. Each currency pair is traded within a separate QCLOB with price-time priority, in which priority is first given to the active orders with the best (i.e., highest buy or lowest sell) price, and ties are broken by selecting the active order with the earliest submission time $t_x$. The platform serves a broad range of trading professionals, including banks, financial institutions, hedge funds, high-frequency traders, corporations, and commodity trading advisers \cite{HotspotWebsite}.\footnote{See \url{http://www.hotspotfx.com/download/userguide/HSFX/HSFX_UserGuide_wrapper.html}.}

\subsection{The Hotspot FX Data}\label{subsec:data}

The data that we study describes all limit order arrivals, cancellations, and trades between $08$:$00$:$00$--$17$:$00$:$00$ GMT for the EUR/USD (Euro/US dollar), GBP/USD (Pounds sterling/US dollar), and EUR/GBP (Euro/Pounds sterling) currency pairs\footnote{A price for the currency pair XXX/YYY denotes how many units of the \emph{counter currency} YYY are exchanged per unit of the \emph{base currency} XXX.} on 30 trading days during May--June 2010. According to the \cite{BIS:2010triennial}, global trade for EUR/USD, GBP/USD, and EUR/GBP constituted about $28\%$, $9\%$, and $3\%$, respectively, of the FX market's total turnover during this period. For each of EUR/USD, GBP/USD, and EUR/GBP, the Hotspot FX platform enforces a minimum order size of $0.01$ units of the base currency and a tick size (i.e., smallest permissible price interval between different orders) of $0.00001$ units of the counter currency.

For each currency pair and each day, the Hotspot FX data consists of two files. The first file is the \emph{tick-data file,} which lists all limit order arrivals and departures and is timestamped to the nearest millisecond. For each limit order arrival, this file lists the price, size, direction (buy/sell), arrival time, and a unique order identifier. For each limit order departure, this file lists the departure time and the departing order's unique identifier. A limit order departure can occur for two reasons: (1) because the order is matched by an incoming market order or (2) because the order is cancelled by its owner. The data provides no way to deduce with certainty whether a given order departure relates to a cancellation or a complete matching.\footnote{When studying order-flow distributions, we treat all active order departures as cancellations. The percentage of active order departures that are actually due to complete matching is extremely low, because market orders constitute about $0.05\%$, about $0.02\%$, and less than $0.01\%$ of arriving order flow for EUR/USD, GBP/USD, and EUR/GBP, respectively (see Table \ref{tab:sfxallagg}). Incorrectly classifying a tiny fraction of departures in this way should have a negligible impact on our results.}

The second file is the \emph{trade-data file,} which lists all trades. For each trade, this file lists the price, size, direction (buy/sell), and trade time, timestamped to the nearest millisecond. If a market order matches to several different active orders, then the trade-data file reports each partial matching as a separate line, with a time stamp that differs from the previous line by at most 1 millisecond. In the absence of explicit details regarding order ownership, we regard all entries that correspond to a trade of the same direction and that arrive within 1 millisecond of each other as originating from the same market order. For each of the three currency pairs, the mean inter-arrival time between trades is of the order of several seconds, so it is unlikely that two separate market orders would arrive within 1 millisecond. We regard any incorrectly grouped market orders as a source of noise in the data.

By processing each order arrival or departure listed in the tick-data file, we are able to reconstruct the global LOB $\mathcal{L}(t)$ at any time during $08$:$00$:$00$--$17$:$00$:$00$ GMT. However, Hotspot FX does not disclose any information regarding CCLs on the platform, and the data contains no information about institutions' identities. Therefore, we are not able to reconstruct any given institution $\theta_i$'s local LOB $\mathcal{L}_i(t)$ from the data. By processing each trade listed in the trade-data file, we are able to reconstruct the trade-price series $B(t)$ and $A(t)$ at any time $t$ during the same period. We are therefore able to calculate both the quote-relative and trade-relative price of any order at any time (see Section \ref{subsec:coordinateframes}).

The data does not provide a reliable way to perform inference about incoming orders that partially match and partially become active. For such orders, we treat the matched part as a market order and the unmatched part as a separate limit order.

\section{Methodology}\label{sec:methodology}

\subsection{Time Scales}

We perform all of our calculations in \emph{event time,} whereby we advance the clock by 1 unit whenever a limit order arrives.\footnote{This includes orders that are only partially filled upon arrival.} Measuring time in this way helps to remove the nonstationarities that occur in calendar time due to irregular bursts of trading activity \cite{Chakraborti:2011empirical, Gourieroux:1999intra, Mantegna:1999introduction, Stephan:1990intraday, Toke:2011market}. The number of market order arrivals and active order cancellations varies in each time unit. We reset the clock at the start of each trading day so that the first limit order arriving after $08$:$00$:$00$ GMT has $t_x=1$.

\subsection{Trading Days}\label{subsec:tradingdays}

To obtain sufficiently many data points to perform statistically stable estimation, some older empirical studies of LOB data aggregate market activity from multiple trading days or multiple different assets \cite{Biais:1995empirical,Bouchaud:2002statistical,Chakraborti:2011empirical,Cont:2010stochastic,Farmer:2005predictive,Gu:2008empiricalregularities,Gu:2008empiricalshape,Mike:2008empirical,Potters:2003more,Zovko:2002power}. However, thanks to increased levels of market activity, technological innovations that facilitate analysis of ever-larger data sets, and improved data quality,\footnote{Several LOB platforms now record data at the accuracy of milliseconds \cite{Hasbrouck:2013low} or even nanoseconds \cite{Bonart:2016latency,Gai:2013externalities,Gould:2016imbalance}. See \cite{Menkveld:2016economics} for a survey of several studies that examine recent LOB data from a wide range of different sources.} aggregating data in this way is less important in empirical studies or more recent LOB data.

Due to the high levels of activity on Hotspot FX and the high quality of the data to which we have access, we are able to study order flow and LOB state on each trading day and for each currency pair separately. We choose a single trading day as our longitudinal unit for three reasons. First, a single trading day represents a structural cycle on Hotspot FX because the platform automatically cancels all active orders at the end of each day \cite{HotspotGUIUserGuide}. Second, a single trading day provides a compromise between including enough data points to ensure statistical stability and including enough longitudinal units to perform useful comparisons. Third, several empirical studies have reported that most institutions implement their investment decisions and trading strategies over a single trading day \cite{Axioglou:2011markets,Bjonnes:2005dealer,Sager:2006under}. To such institutions, statistics that describe market behaviour over this time horizon are likely to be the most useful.

\subsection{Buy and Sell Orders}

The use of quote-relative and trade-relative coordinates facilitates the aggregation of buy and sell orders into a single data set (see Section \ref{subsec:coordinateframes}). Throughout this paper, we report all of our results for buy and sell orders together, because aggregating the data in this way increases the sample size when compared to studying buy or sell orders separately. We repeated all of our calculations for buy and sell orders separately, and we obtained qualitatively similar results to those that we report, albeit with a smaller sample size and a correspondingly larger statistical noise.

\section{Results}\label{sec:results}

\subsection{LOB Activity}

In Table \ref{tab:sfxallagg}, we list summary statistics that describe aggregate LOB activity for EUR/USD, GBP/USD, and EUR/GBP on Hotspot FX across all 30 trading days in our sample. In terms of both limit order and market order arrivals, EUR/USD is the most active and EUR/GBP is the least active of the three currency pairs. The total volume of arriving limit orders is about $30\%$ larger for GBP/USD and about $60\%$ larger for EUR/USD than it is for EUR/GBP. The corresponding results for market orders are even more extreme: the total size of market order arrivals for GBP/USD and EUR/USD outstrip that of EUR/GBP by a factor of about 4 and a factor of more than 10, respectively. Therefore, comparing our subsequent results for the three different currency pairs enables us to contrast the behaviour of the QCLOBs for currency pairs with substantially different levels of trading activity.

\begin{table}[!htbp]
\footnotesize
\begin{center}
\begin{tabular}{|ll|l|l|l|}
\hline
 & & EUR/USD & GBP/USD & EUR/GBP \\
\hline
\multirow{3}{*}{Total size (units of base currency $\times 10^{9}$)} & Limit orders & $301136$ & $235934$ & $184597$ \\ 
 & Market orders & $137$ & $46$ & $12$ \\ 
 & Cancellations & $300959$ & $235868$ & $184580$ \\ 
\hline
\multirow{3}{*}{Total number (orders $\times 10^{3}$)}  & Limit orders & $136009$ & $131088$ & $87982$ \\ 
 & Market orders & $168$ & $87$ & $15$ \\ 
 & Cancellations & $135805$ & $130987$ & $87964$ \\ 
\hline
\multirow{3}{*}{Mean inter-arrival time (seconds)} & Limit orders & $0.00715$ & $0.00741$ & $0.011$ \\ 
 & Market orders & $5.78$ & $11.1$ & $62.9$ \\ 
 & Cancellations & $0.00716$ & $0.00742$ & $0.011$ \\ 
\hline
\multirow{3}{*}{Modal size (units of base currency $\times 10^{6}$)} & Limit orders & $1.00$ & $1.00$ & $1.00$ \\ 
 & Market orders & $1.00$ & $1.00$ & $1.00$ \\ 
 & Cancellations & $1.00$ & $1.00$ & $1.00$ \\ 
\hline
\multirow{3}{*}{Mean size (units of base currency $\times 10^{6}$)} & Limit orders & $2.21$ & $ 1.8$ & $ 2.1$ \\ 
 & Market orders & $0.818$ & $0.523$ & $0.777$ \\ 
 & Cancellations & $2.22$ & $ 1.8$ & $ 2.1$ \\ 
\hline
 \multicolumn{2}{|l|}{Percentage of market orders that match at several different prices} & $8.41\%$ & $ 6.3\%$ & $4.25\%$ \\ 
\hline
 \multicolumn{2}{|l|}{Mean total size of active orders (units of base currency $\times 10^{6}$)} & $579$ & $330$ & $189$ \\ 
\hline
 \multicolumn{2}{|l|}{Mean total depth at best quotes (units of base currency $\times 10^{6}$)} & $6.04$ & $ 4.8$ & $4.97$ \\ 
\hline
\end{tabular}
\caption{Summary statistics for aggregate activity on all 30 trading days that we study.}
\label{tab:sfxallagg}
\end{center}
\end{table}

For each of the three currency pairs, limit order arrivals outstrip market order arrivals by more than 3 orders of magnitude. Market orders constitute less than $0.05\%$ of the total arriving order flow, which indicates that the vast majority of limit orders end in cancellation rather than matching. Indeed, in each case, the total size of cancellations is very close to the total size of limit order arrivals. The remaining volume of limit orders (not accounted for either by matching or by cancellation) indicates that the mean total size of active orders in the global LOBs increases on average through the trading day.

\begin{figure}[!htbp]
\centering
\includegraphics[width=0.8\textwidth]{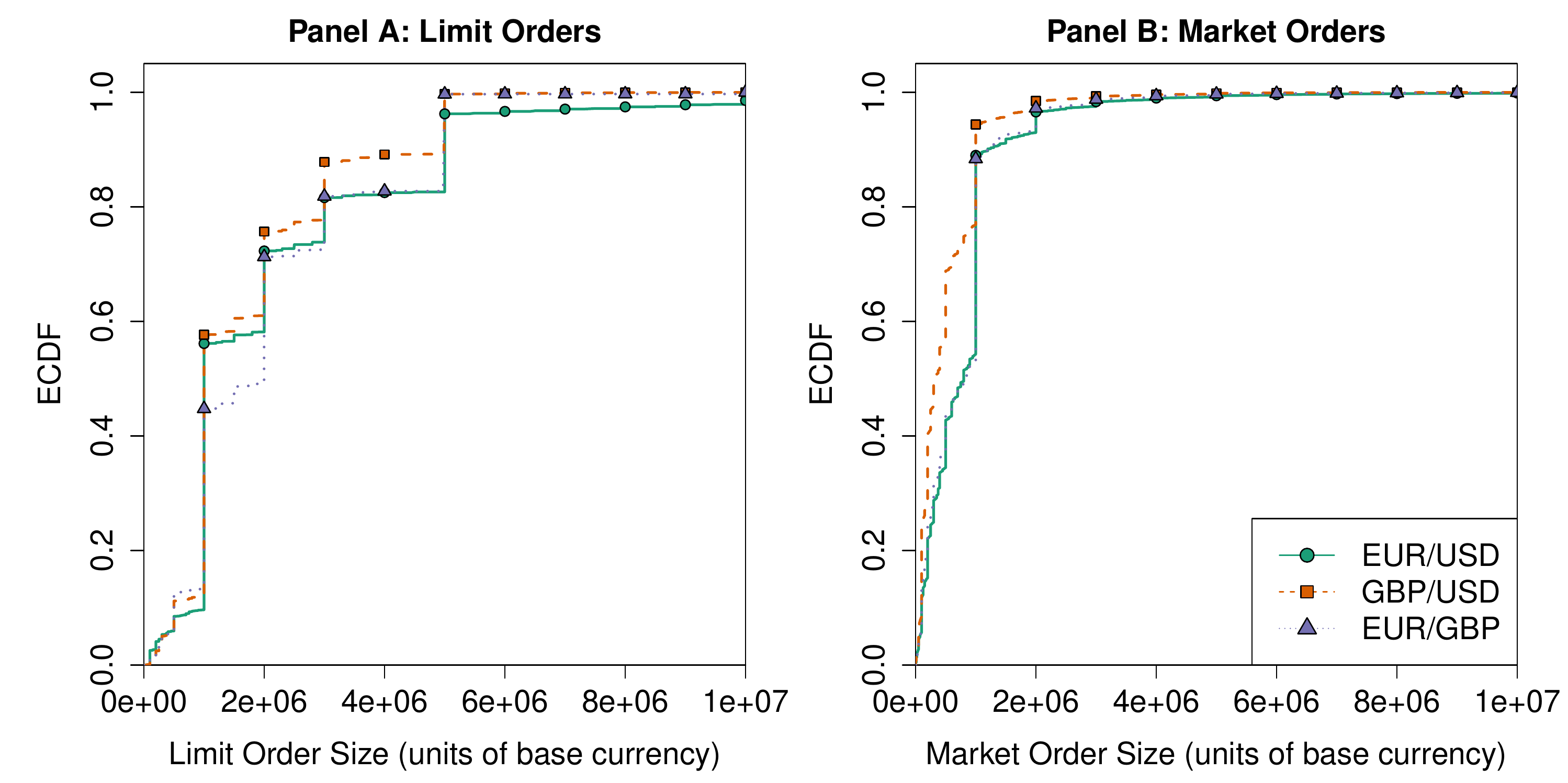}
\caption{Empirical cumulative density functions (ECDFs) of the sizes of arriving (Panel A) limit orders and (Panel B) market orders for (solid green curves with circles) EUR/USD, (dashed orange curves with squares) GBP/USD, and (dotted purple curves with triangles) EUR/GBP.}
\label{fig:sfxLOMOecdf}
\end{figure}

For both limit orders and market orders, the modal size is exactly 1 million units of the base currency. The empirical cumulative density functions (ECDFs) of order sizes (see Figure \ref{fig:sfxLOMOecdf}) reveal that institutions favour orders with round-number sizes that are integer multiples of 1 million, even though the minimum order size on Hotspot FX is just $0.01$ units of the base currency (see Section \ref{sec:hotspotdata}). Despite their common mode, the mean size of arriving limit orders for each currency pair is more than double the corresponding number for market orders due to the higher concentration of small market order sizes than of small limit order sizes.

For each of the three currency pairs, the mean total depth at the best quotes (i.e., the mean total size of active orders at $b(t)$ or $a(t)$) is less than $1\%$ of the total size of all active orders. Despite this relatively small fraction of liquidity at the global best prices, it still exceeds the mean size of market orders by a factor of almost 10 in each case. Moreover, only a small percentage of market orders match at more than one price. Together, these results suggest that institutions employ \emph{selective liquidity-taking,} in the sense that they carefully monitor the market state to ensure that they only conduct trades at favourable prices.\footnote{For a detailed introduction to selective liquidity-taking, see \cite{Bouchaud:2009digest}.}

\begin{figure}[!htbp]
\centering
\includegraphics[width=0.8\textwidth]{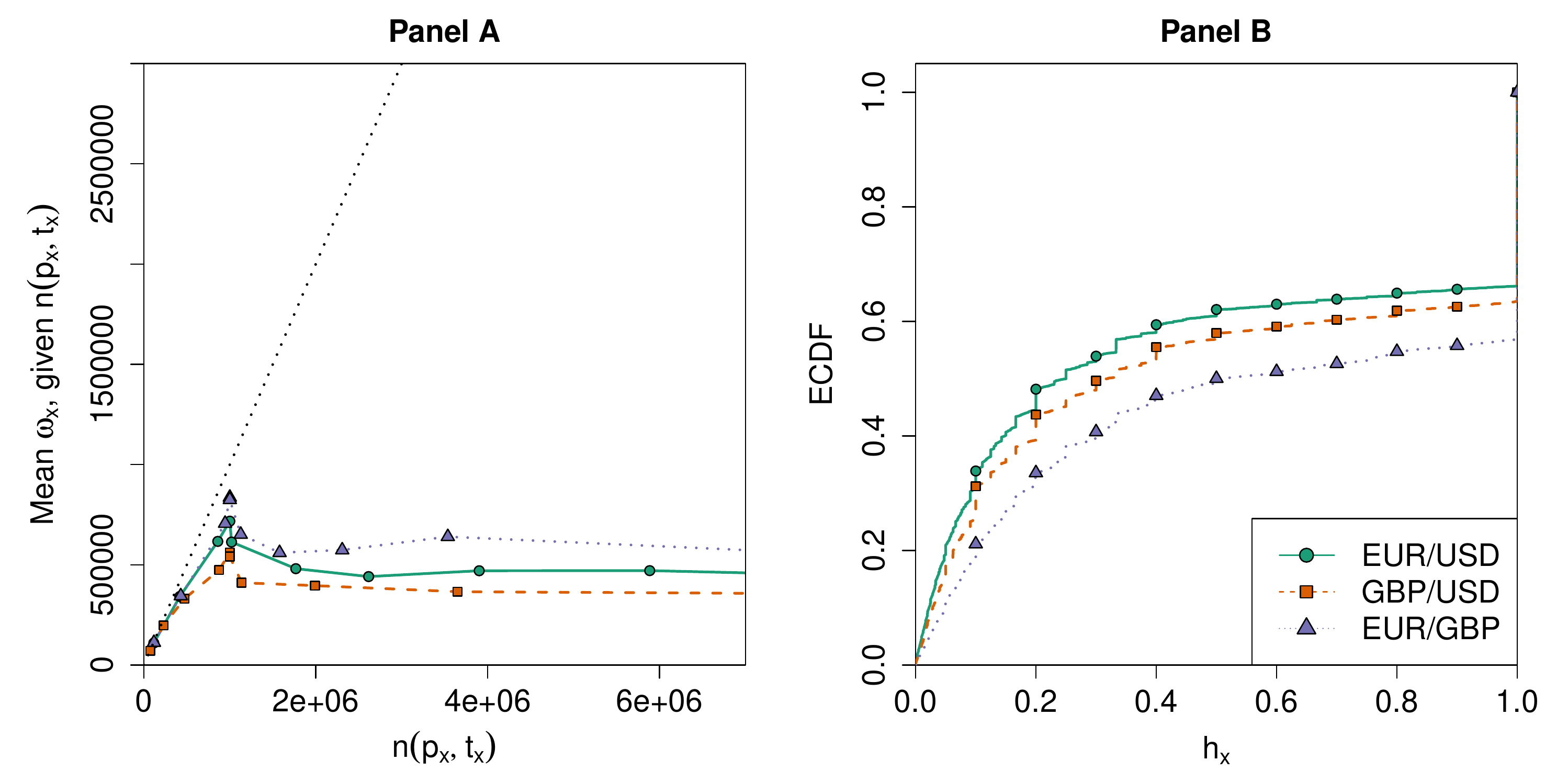}
\caption{(Panel A) Mean order size among market orders that match to a queue of a given length and (Panel B) ECDFs of the fraction of the queue depth consumed by an incoming market order for (solid green curve with circles) EUR/USD, (dashed orange curve with squares) GBP/USD, and (dotted purple curve with triangles) EUR/GBP. In Panel A, we bin the data into deciles according to queue length. The dotted black line in Panel A indicates the diagonal.}
\label{fig:sfxLOMOfraction}
\end{figure}

We now assess the relationship between the sizes of market orders and the sizes of the queues to which they match. In Panel A of Figure \ref{fig:sfxLOMOfraction}, we show how the mean order size varies among market orders that match to a queue of a given length. For all queue lengths, the mean size of arriving market orders is strictly smaller than the queue length. This result is consistent with our observation that it is relatively rare for market orders to match at more than one price. For queue lengths up to about 1 million, the mean size of market orders grows approximately linearly with the queue length, with a scale factor that varies across the three currency pairs but is less than 1 in each case. However, this does not persist for queue sizes longer than about 1 million, for which the mean market order size becomes approximately constant for each of the three currency pairs. This finding contrasts to the results reported by \cite{Farmer:2004what} for order flow on the LSE (which operates as a centralized LOB), in which the approximately linear relationship that we observe for small queue lengths persists across the whole domain, even when the total depth of active orders at the best quotes is very large. In Section \ref{sec:sfxdiscussion}, we return to this discussion and propose two possible explanations for the behaviour that we observe on Hotspot FX.

To further illustrate the presence of selective liquidity-taking, we also calculate the fraction of the relevant queue depth that each market order consumes upon arrival. For a sell market order $x$ submitted at time $t_x$ with price $p_x$ and size $\omega_x>0$, we calculate the ratio$$h_x = \left| \frac{\omega_x}{n^b(p_x,t_x)} \right|,$$where $n^b(p_x,t_x)$ denotes the total size of active buy orders in the global LOB with price $p_x$ immediately before the market order arrival at time $t_x$. For a buy market order $x$, we calculate the same ratio $h_x$, but we use the corresponding total size $n^a(p_x,t_x)$ of active sell orders. In Panel B of Figure \ref{fig:sfxLOMOfraction}, we show the ECDFs of $h_x$.

Our results paint an interesting picture of selective liquidity-taking on Hotspot FX. On the one hand, about $33\%$ of market orders for EUR/USD, about $36\%$ of market orders for GBP/USD, and about $43\%$ of market orders for EUR/GBP consume the entire queue to which they match. This suggests that a considerable fraction of institutions condition their market order size to match the depth of active orders available. On the other hand, some market orders consume a relatively small fraction of the relevant queue depth. For example, about half of all market orders for EUR/USD consume less than $20\%$ of the relevant queue depth. This may indicate that the institutions that submit these market orders do not wish to perform large trades, despite large depths being available to them. However, it may also be the case that these institutions do not have sufficient CCLs to access the full depths available in the global LOB, and that they therefore instead condition their market order sizes to the depth available in their local LOB. We also return to this discussion in Section \ref{sec:sfxdiscussion}.

In Table \ref{tab:sfxspread}, we list summary statistics for the global bid--ask spread $s(t)$. Both the mean and median values of $s(t)$ are similar for GBP/USD and EUR/GBP, but they are much smaller for EUR/USD. This implies that $s(t)$ tends to be smaller for EUR/USD than for the other two currency pairs. In a centralized LOB, a smaller value of $s(t)$ is often construed as a sign of greater liquidity \cite{Ding:2010electronic}, because $s(t)$ determines the cost of conducting a round-trip trade (i.e., buying a single unit at $a(t)$ and selling a single unit at $b(t)$ using a pair of simultaneous market orders). In a QCLOB, by contrast, $s(t)$ does not have such a clear interpretation because the liquidity available to each institution $\theta_i$ depends on its local LOB $\mathcal{L}_i(t)$.

\begin{table}[!htbp]
\footnotesize
\begin{center}
\begin{tabular}{|l|l|l|l|}
\hline
 & EUR/USD & GBP/USD & EUR/GBP \\
\hline
Minimum (ticks) & $-365$ & $-270$ & $-60$ \\ 
\hline
Maximum (ticks) & $69$ & $147$ & $152$ \\ 
\hline
Median (ticks) & $4$ & $10$ & $10$ \\ 
\hline
Mean (ticks) & $3.62$ & $9.54$ & $10.11$ \\ 
\hline
Percentage of time for which $s(t)<0$ & $9.99\%$ & $4.08\%$ & $0.23\%$ \\ 
\hline
Mean duration for which $s(t)<0$ (seconds) & $0.10$ & $0.12$ & $0.16$ \\ 
\hline
Mean crossed volume (units of base currency $\times 10^{6}$) & $9.50$ & $7.61$ & $5.11$ \\ 
\hline
\end{tabular}
\caption{Summary statistics for the global bid--ask spread $s(t)$.}
\label{tab:sfxspread}
\end{center}
\end{table}

Another important contrast between centralized LOBs and QCLOBs is that the global spread $s(t)$ is always strictly positive in a centralized LOB, but can become negative in a QCLOB (see Figure \ref{fig:unfilteredfiltered}). This occurs whenever there exist a buy limit order $x$ and a sell limit order $y$ such that $p_y < p_x$. In a centralized LOB, the arrival of the second such order would trigger an immediate matching, so $x$ and $y$ would never coexist in $\mathcal{L}(t)$. In a QCLOB, however, if the CCLs between the institutions that own $x$ and $y$ do not permit them to perform the corresponding trade, then both $x$ and $y$ can be active simultaneously. Therefore, the global bid--ask spread can be negative in a QCLOB. However, as discussed in Section~\ref{subsec:qclobs}, negative spreads need not indicate the existence of tradable arbitrage opportunities, because such opportunities may not be permitted by the CCL structure.

\begin{figure}[!htbp]
\centering
\includegraphics[width=0.8\textwidth]{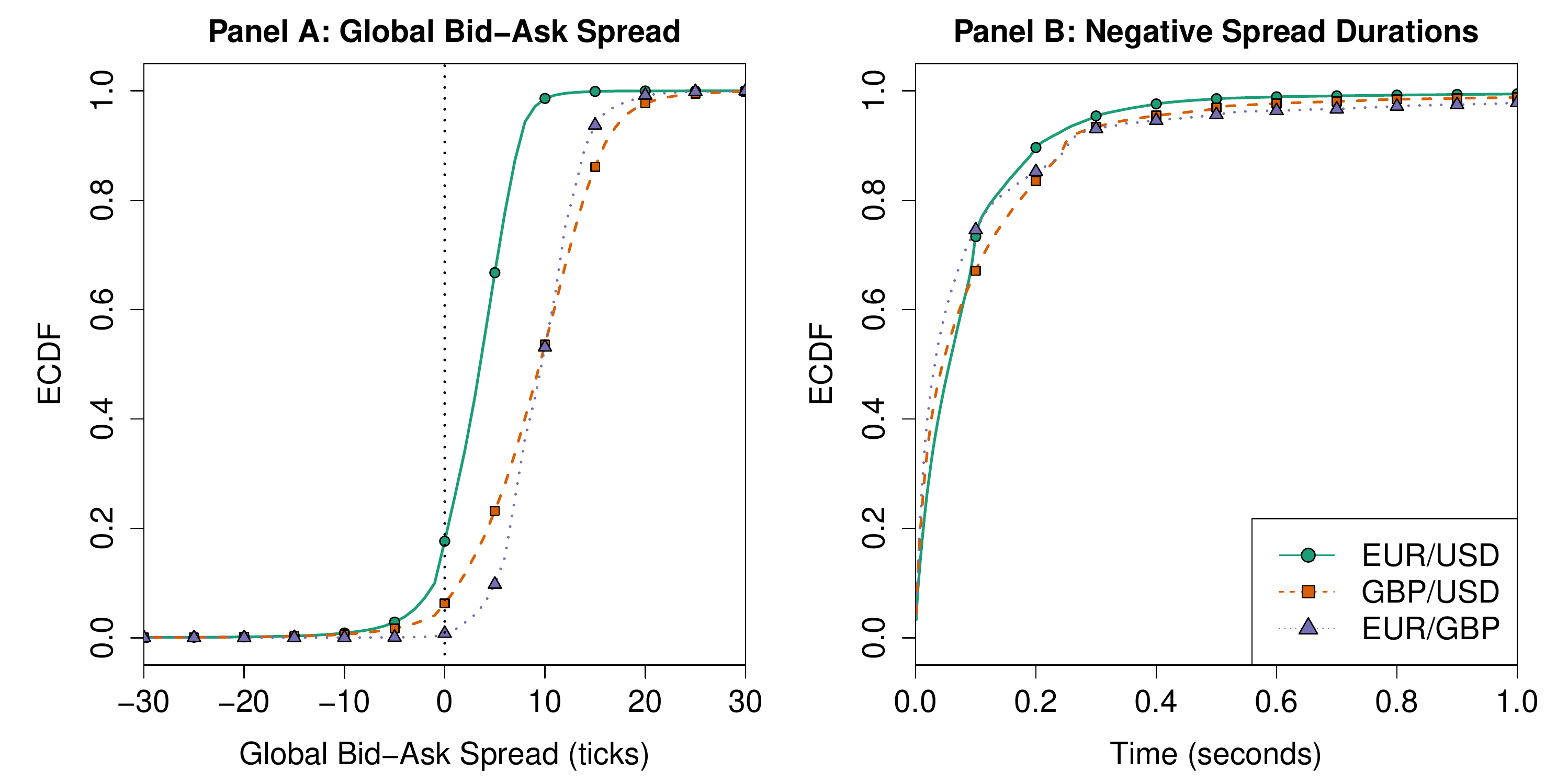}
\caption{ECDFs of (Panel A) the global bid--ask spread $s(t)$ and (Panel B) the duration, measured in seconds, for which $s(t)$ remains negative, for (solid green curve with circles) EUR/USD, (dashed orange curve with squares) GBP/USD, and (dotted purple curve with triangles) EUR/GBP. The dotted black line in Panel A indicates $s(t)=0$.}
\label{fig:sfxSpread}
\end{figure}

In Panel A of Figure \ref{fig:sfxSpread}, we show the ECDF of $s(t)$. As we also illustrate in Table \ref{tab:sfxspread}, the global bid--ask spread is negative for almost $10\%$ of the time for EUR/USD and for more than $4\%$ of the time for GBP/USD, but it is rarely negative for EUR/GBP. In the most extreme case (which occurs for EUR/USD), the spread is more than $350$ ticks negative. Among the times when $s(t)$ is negative, the mean crossed volume (i.e., the total size of all sell orders with $p_x<b(t)$ and all buy orders with $p(x)>a(t)$) is about $10$ million for EUR/USD, about $7.5$ million for GBP/USD, and about $5$ million for EUR/GBP. In Panel B of Figure \ref{fig:sfxSpread}, we show the ECDF of time durations for which $s(t)$ remains negative (i.e., the ECDF of time differences between when the spread becomes negative and when it next becomes positive). The global bid--ask spread typically remains negative for extremely short durations.

\subsection{Daily Activity Levels}

\begin{figure}[!htbp]
\centering
\includegraphics[width=0.8\textwidth]{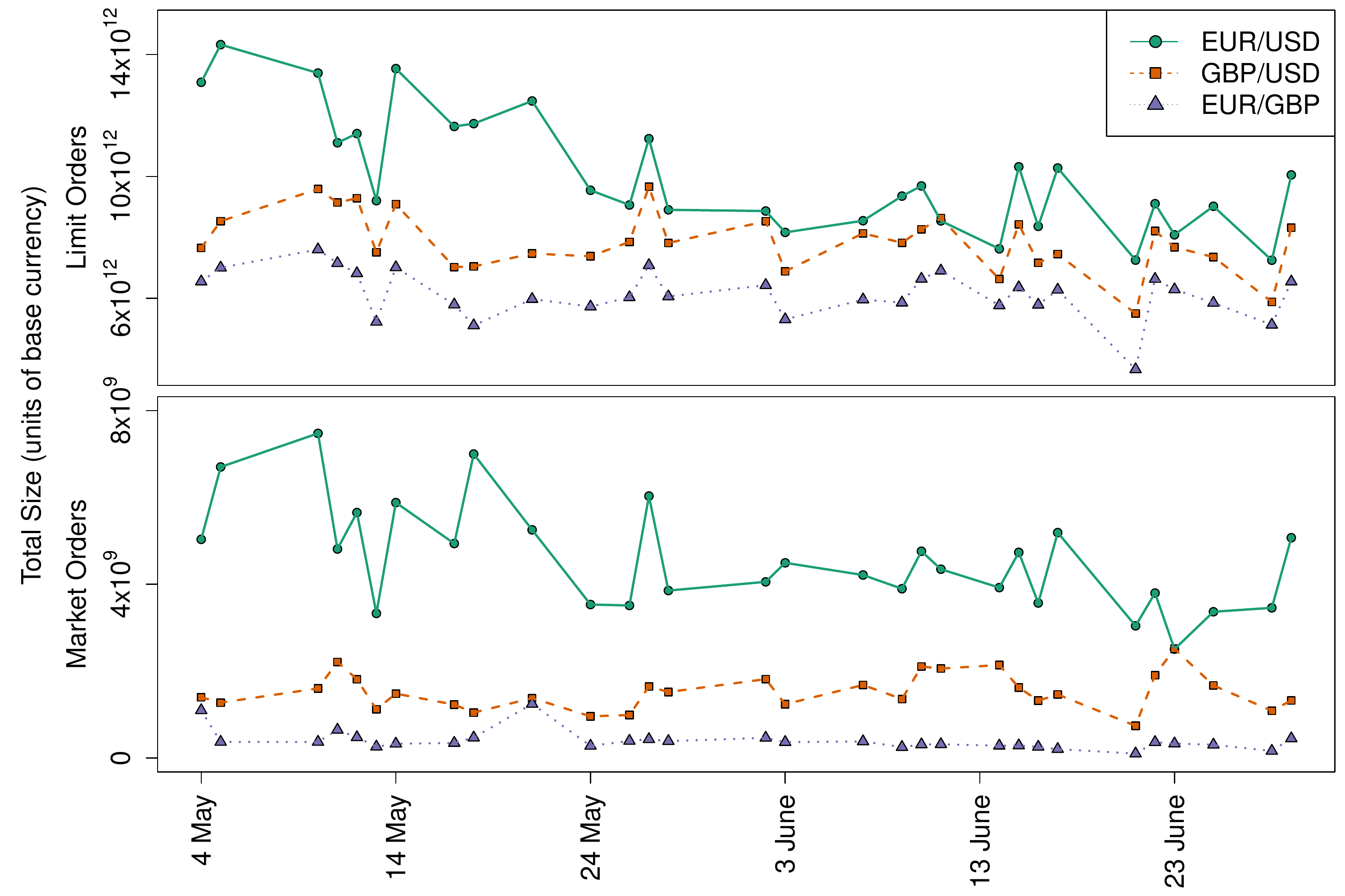}
\caption{Total size of arriving (top panel) limit orders and (bottom panel) market orders for (solid green curve with circles) EUR/USD, (dashed orange curve with squares) GBP/USD, and (dotted purple curve with triangles) EUR/GBP.}
\label{fig:sfxtotalsizelomocx}
\end{figure}

In Figure \ref{fig:sfxtotalsizelomocx}, we show the total size of arriving limit orders and market orders on each of the 30 days in our sample. Although aggregate market activity levels vary considerably across trading days, especially active or especially quiet days tend to coincide for each of the three currency pairs (particularly for limit order arrivals). This suggests that common, exogenous factors play an important role in institutions' trading decisions. In May 2010, the European Central Bank announced and implemented a series of measures to combat financial instability within the Eurozone; these included providing loans to countries in financial difficulties, recapitalizing financial institutions, and purchasing bonds from member states \cite{EFSFFAQ,EFSFWebsite}. The large changes in daily aggregate activity levels during May 2010 suggest that the implementation of such measures and the uncertainty surrounding their announcements strongly influenced activity in the FX spot market.

\subsection{Activity on a Single Trading Day}\label{subsec:sfxlodist}

We next calculate these distributions in a single trading day, to help understand the distributions of order flow and LOB state across different quote- and trade-relative prices. We arbitrarily choose to present the results for 4 May 2010, which is the first day in our sample. In Section \ref{subsec:differentdays}, we investigate how these distributions vary across trading days.

\begin{figure}[!htbp]
\centering
\includegraphics[width=0.8\textwidth]{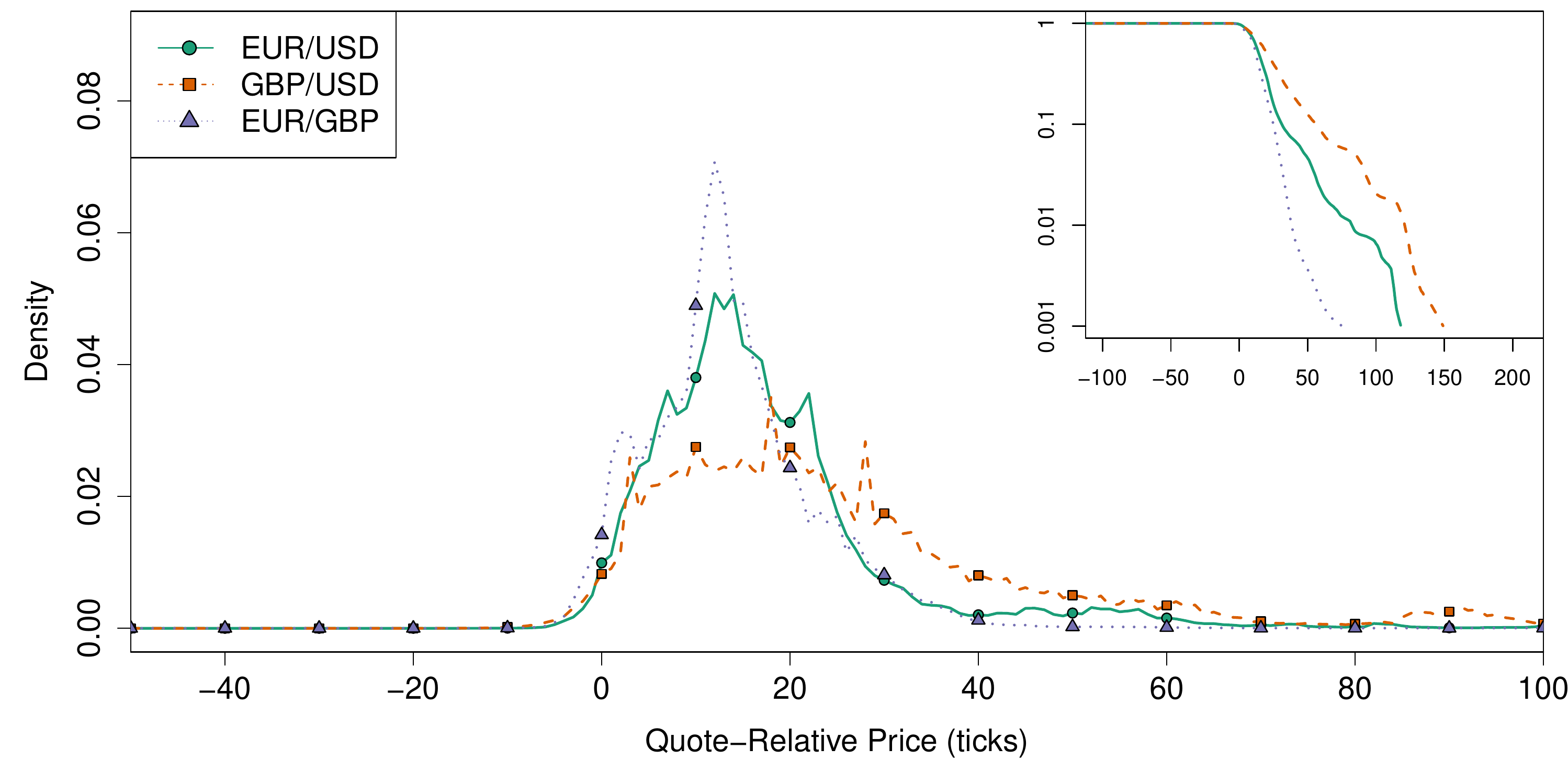}
\includegraphics[width=0.8\textwidth]{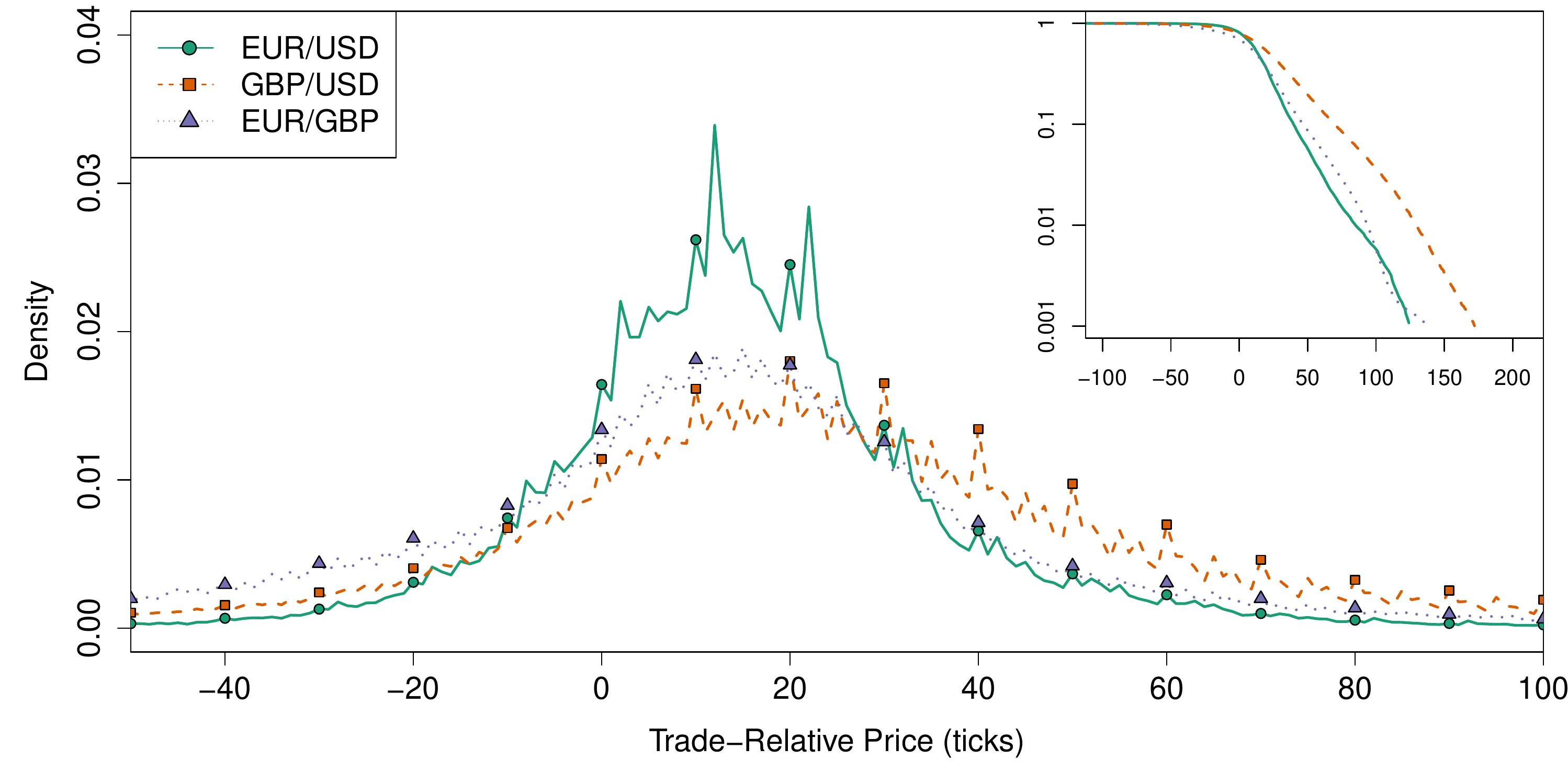}
\caption{Distributions of limit order arrivals for (solid green curves with circles) EUR/USD, (dashed orange curves with squares) GBP/USD, and (dotted purple curves with triangles) EUR/GBP on 4 May 2010 in (top) quote-relative and (bottom) trade-relative coordinates. The main plots show the empirical density functions, and the inset plots show the corresponding survivor functions (i.e., $1-F(x)$, where $F$ is the ECDF) in semi-logarithmic coordinates.}
\label{fig:sfxLOdistday}
\end{figure}

In Figure \ref{fig:sfxLOdistday}, we show the quote-relative and trade-relative price distributions of limit order arrivals on 4 May 2010. For each of the three currency pairs, the maximum limit order arrival rate occurs at a strictly positive relative price in both quote-relative (12 ticks for EUR/USD, 18 ticks for GBP/USD, and 14 ticks for EUR/GBP) and trade-relative (12 ticks for EUR/USD, 20 ticks for GBP/USD, and 15 ticks for EUR/GBP) coordinates. Some institutions place limit orders with extremely large quote- and trade-relative prices, which suggests that they seek to profit from large price swings on long time horizons.

\begin{figure}[!htbp]
\centering
\includegraphics[width=0.8\textwidth]{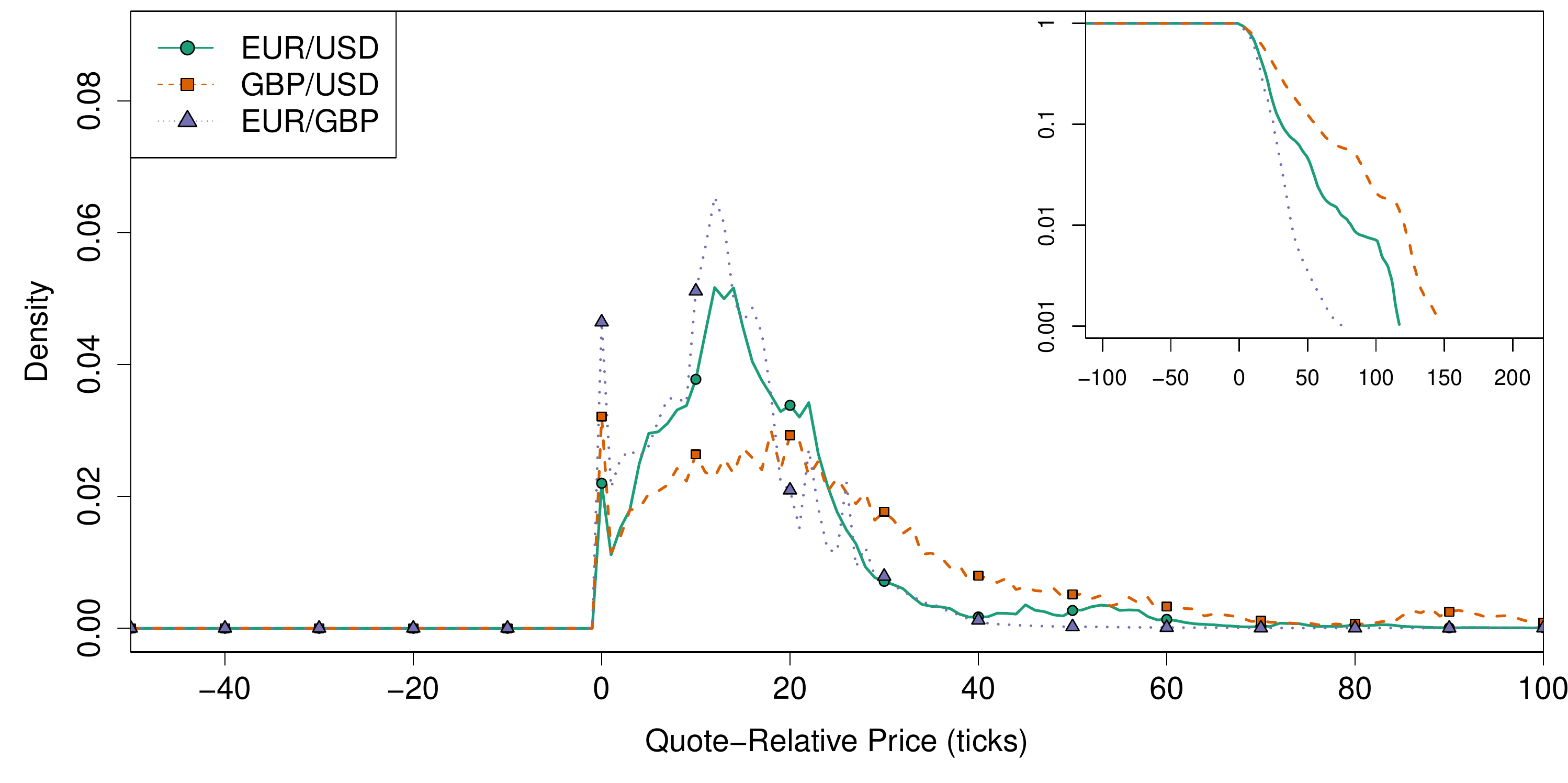}
\includegraphics[width=0.8\textwidth]{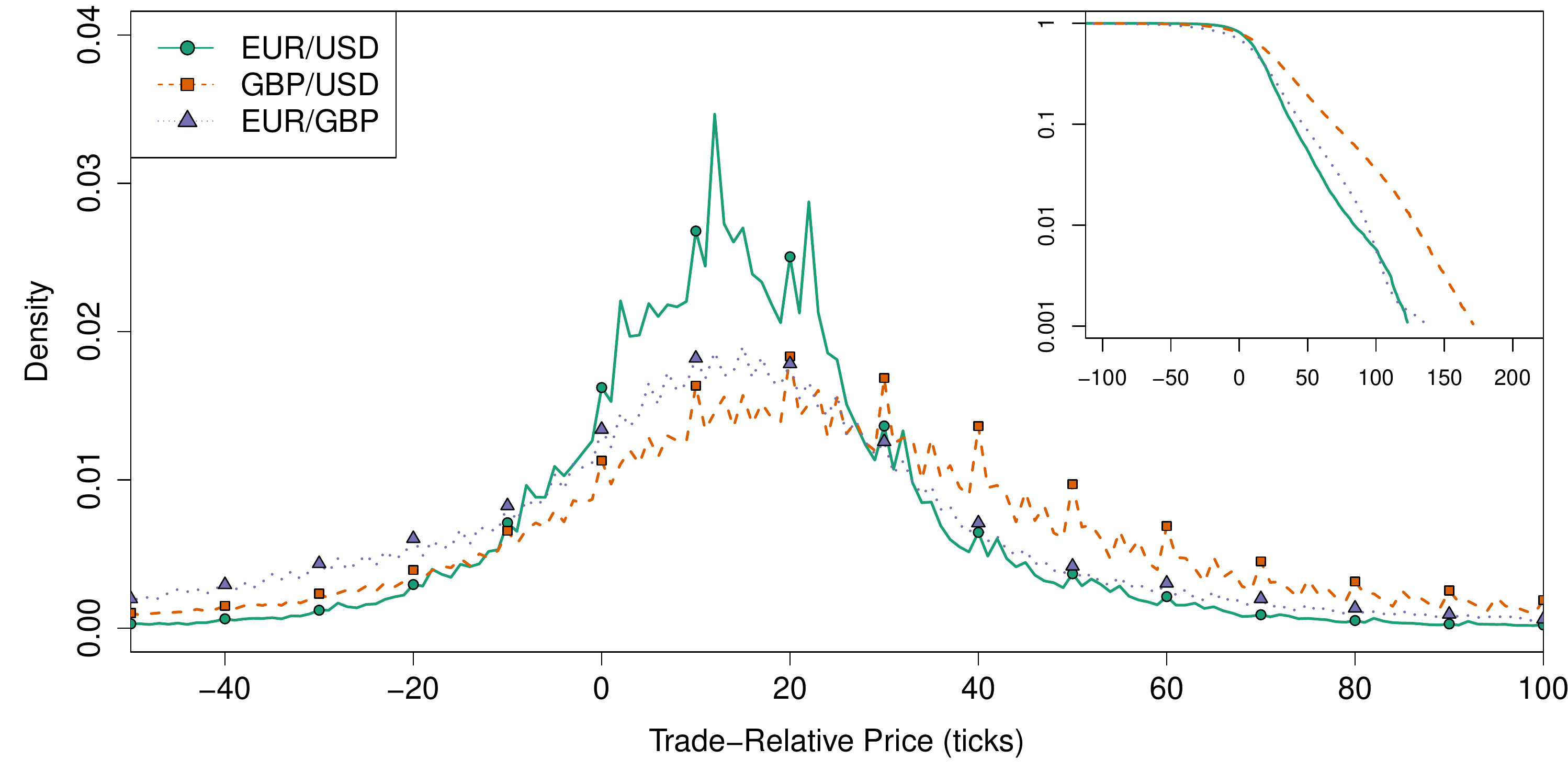}
\caption{Distributions of cancellations by (top) quote-relative and (bottom) trade-relative price for (solid green curves with circles) EUR/USD, (dashed orange curves with squares) GBP/USD, and (dotted purple curves with triangles) EUR/GBP on 4 May 2010 in (top) quote-relative and (bottom) trade-relative coordinates. The main plots show the empirical density functions, and the inset plots show the corresponding survivor functions (i.e., $1-F(x)$, where $F$ is the ECDF) in semi-logarithmic coordinates.}
\label{fig:sfxCXdistday}
\end{figure}

In Figure \ref{fig:sfxCXdistday}, we show the quote-relative and trade-relative distributions of cancellations for each of the three currency pairs. In contrast to limit order arrivals, cancellations can only occur at non-negative quote-relative prices, because the lowest possible quote-relative price of an active order is 0 (which occurs for orders at $b(t)$ or $a(t)$). Each of the three currency pairs' quote-relative cancellation distributions have a local maximum at 0. Cancellations for GBP/USD tend to occur further from the best quotes and cancellations for EUR/USD tend to occur closer to the most recently traded price than do those for the other two currency pairs. For strictly positive quote-relative prices, the cancellation distributions have qualitatively similar shapes to the corresponding distributions for limit order arrivals. In trade-relative coordinates, the cancellation distributions are extremely similar to the corresponding limit order arrival distributions at all prices.

\begin{figure}[!htbp]
\centering
\includegraphics[width=0.8\textwidth]{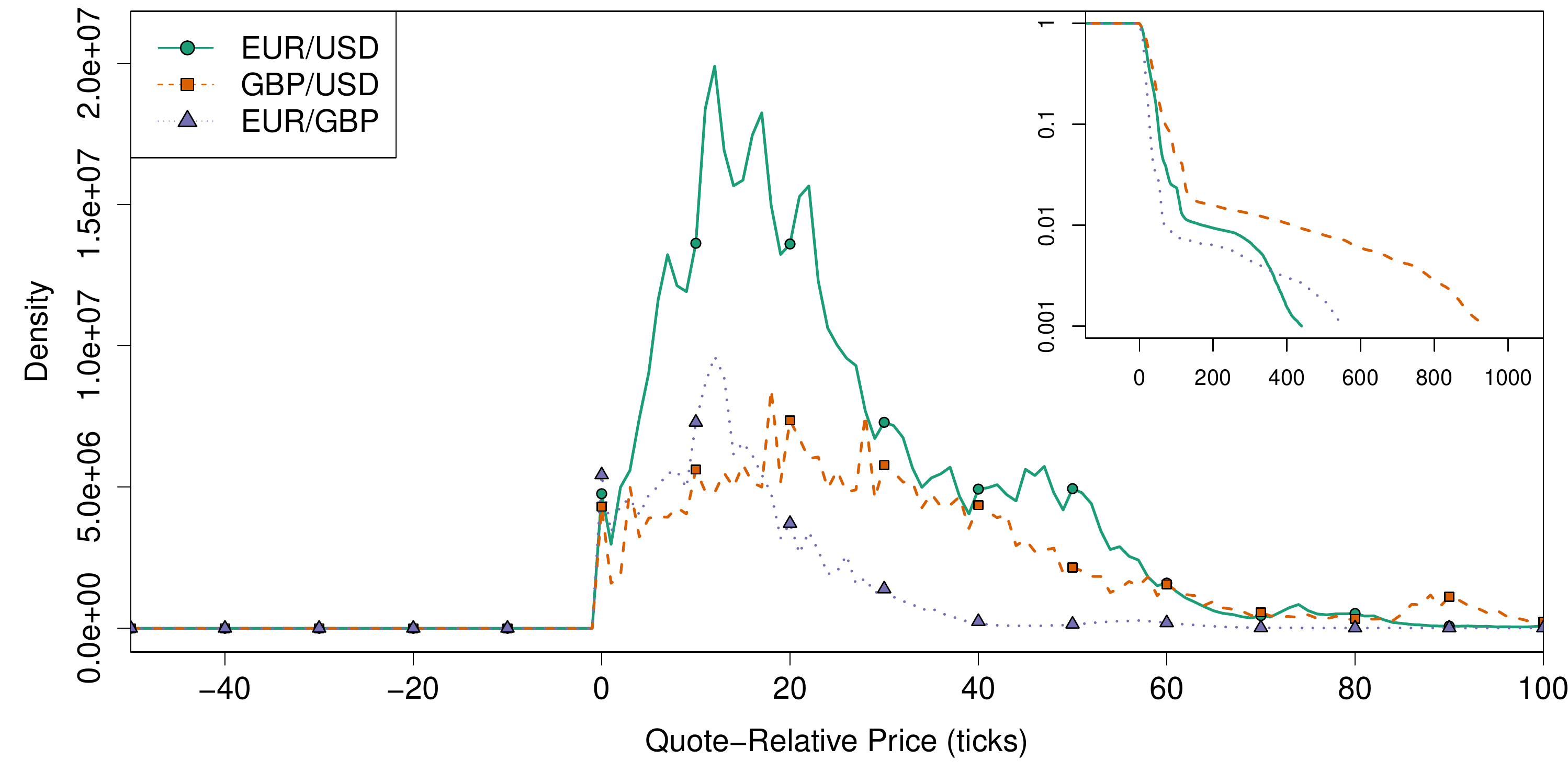}
\includegraphics[width=0.8\textwidth]{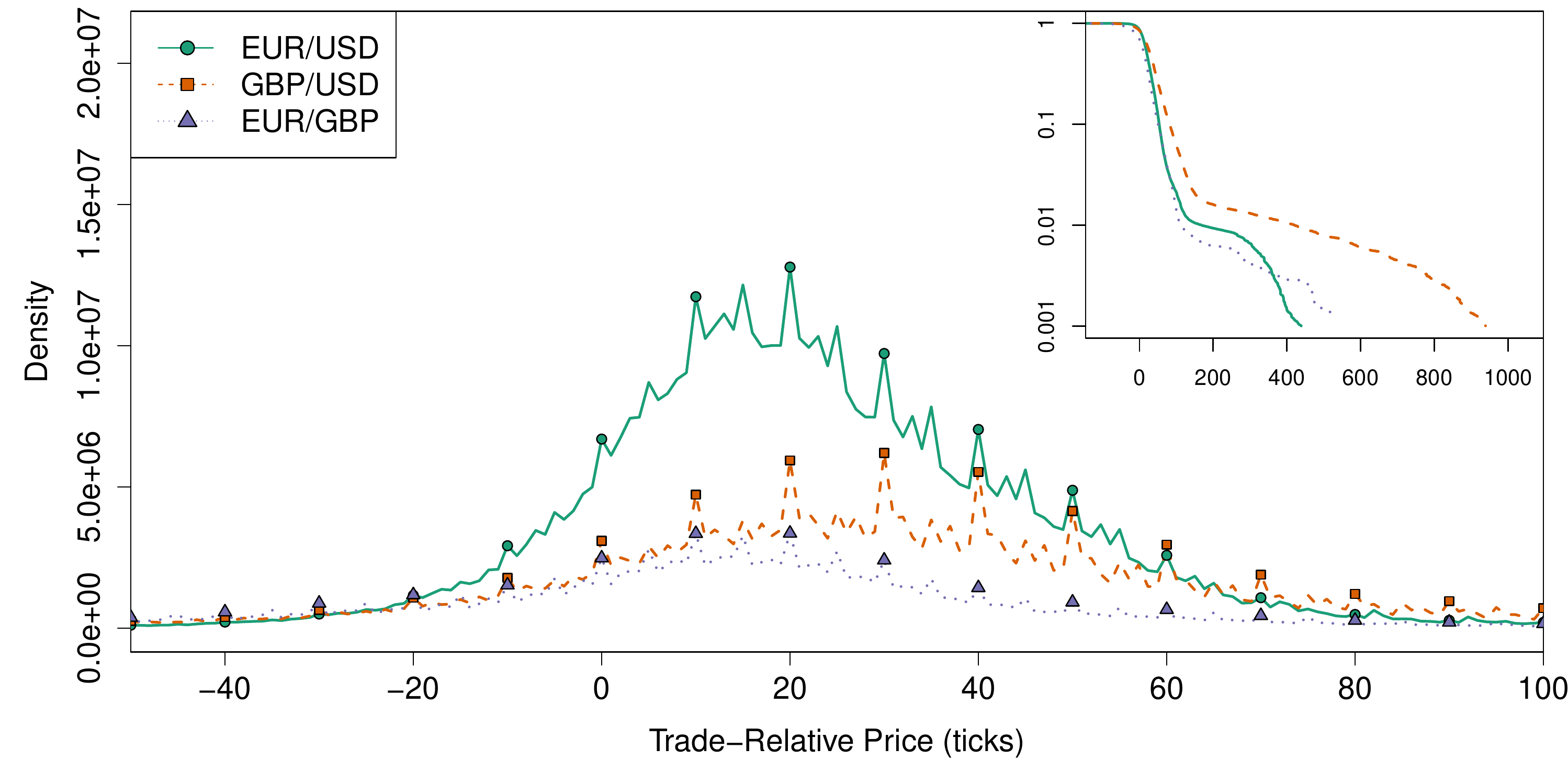}
\caption{Mean depths at given (top) quote-relative and (bottom) trade-relative prices for (solid green curves with circles) EUR/USD, (dashed orange curves with squares) GBP/USD, and (dotted purple curves with triangles) EUR/GBP on 4 May 2010. The plots show the total absolute size of both buy and sell orders at the given relative prices. The main plots show the empirical mean depths (in units of the base currency), and the inset plots show the normalized empirical cumulative mean depths (i.e., the empirical cumulative mean depths expressed as a fraction of the mean total size of all active orders at all prices).}
\label{fig:sfxDPdistday}
\end{figure}

In Figure \ref{fig:sfxDPdistday}, we show the mean depths (i.e., the mean total size of active orders in the global LOB $\mathcal{L}(t)$) at given quote-relative and trade-relative prices. By definition, the mean depth is 0 for all negative quote-relative prices. Although all three currency pairs have a local maximum in mean depth at the best quotes, in each case, it is much smaller than the corresponding local maximum in the cancellation distributions. In both quote-relative and trade-relative coordinates, the mean depth at small quote-relative prices is substantially larger for EUR/USD than it is for GBP/USD and EUR/GBP. In trade-relative coordinates, each currency pair's local maximum occurs at a strictly positive relative price (20 ticks for EUR/USD and EUR/GBP, and 30 ticks for GBP/USD). The upper tails of the distribution of mean depths are much heavier than those of the corresponding distributions of limit order arrivals and cancellations. This suggests that some institutions leave active orders far from the best quotes for long periods of time. Although such orders constitute a tiny fraction of the aggregate order flow, their long lifetimes cause them to contribute significantly to the mean depths when averaged across the whole sample period.

Figures \ref{fig:sfxLOdistday} and \ref{fig:sfxCXdistday} illustrate an interesting round-number effect in order flow: limit order arrivals and cancellations occur more frequently at relative prices that are integer multiples of 10 than they do at neighbouring relative prices. Similarly, Figure \ref{fig:sfxDPdistday} illustrates that the total depth in $\mathcal{L}(t)$ tends to be larger at relative prices that are integer multiples of 10 than it does at neighbouring relative prices.

\begin{figure}[!htbp]
\centering
\includegraphics[width=0.8\textwidth]{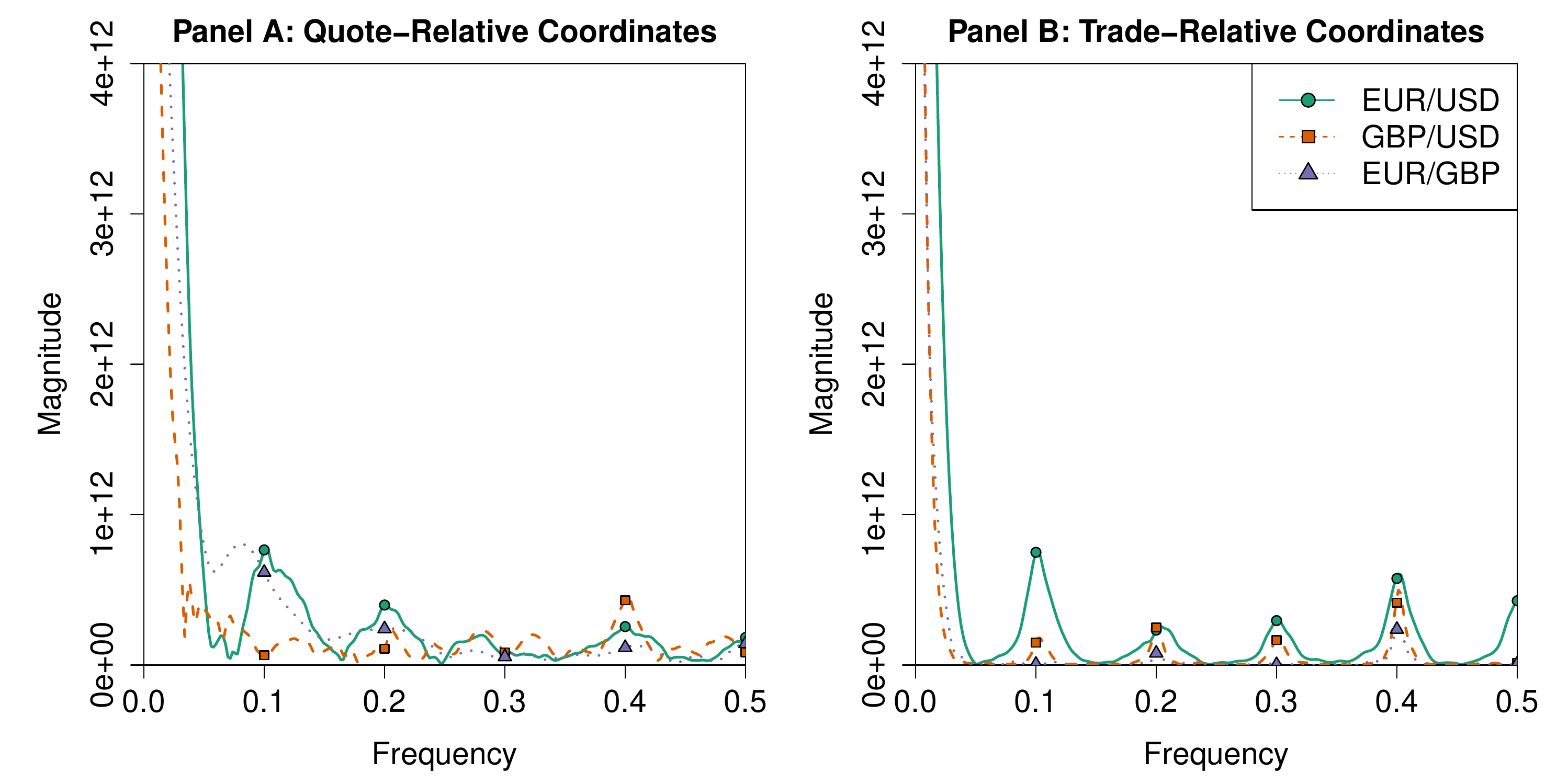}
\caption{Magnitude spectra for the distributions of limit order arrivals for (solid green curves with circles) EUR/USD, (dashed orange curves with squares) GBP/USD, and (dotted purple curves with triangles) EUR/GBP on 4 May 2010 in (Panel A) quote-relative and (Panel B) trade-relative coordinates. We obtain the magnitudes by calculating the absolute value of the fast Fourier Transform (FFT) of the corresponding empirical density functions from Figure \ref{fig:sfxLOdistday}.}
\label{fig:sfxFFT}
\end{figure}

To help quantify the strength of this effect, we calculate magnitude spectra by applying the fast Fourier Transform (FFT) to the corresponding empirical density functions. In Figure \ref{fig:sfxFFT}, we show the magnitude spectra of limit order arrivals on 4 May 2010. The corresponding plots for cancellations and mean depths are qualitatively similar (however, they are slightly noisier). In quote-relative coordinates, the magnitude spectra exhibit a weak periodicity at integer multiples of $0.1$ (which corresponds to a period of 10 ticks), but they also contain several other local maxima close to these peaks. In trade-relative coordinates, the magnitude spectra exhibit a much stronger signature of periodicity at integer multiples of $0.1$, with clear local maxima corresponding to these frequencies.

\begin{figure}[!htbp]
\centering
\includegraphics[width=0.8\textwidth]{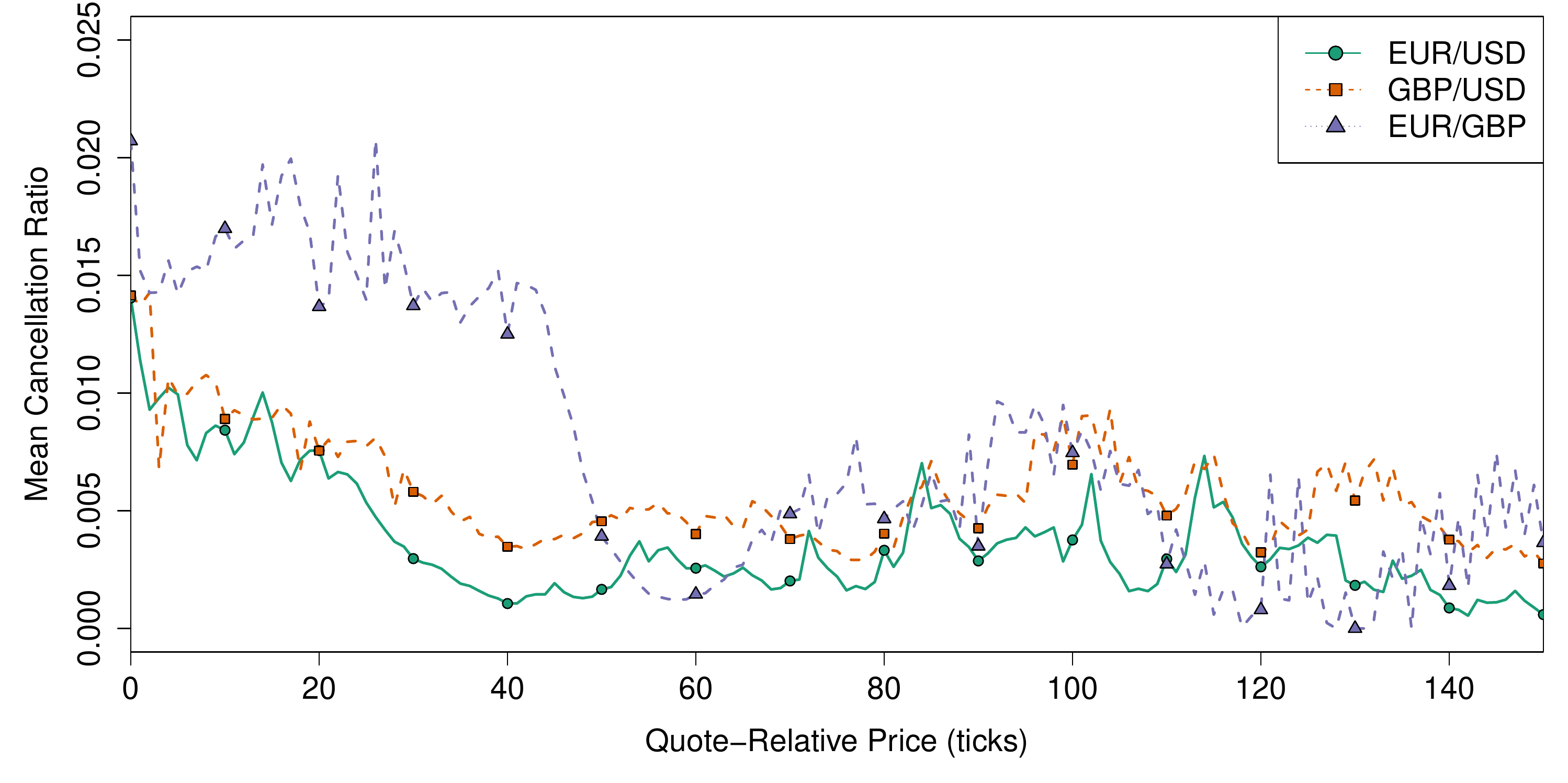}
\includegraphics[width=0.8\textwidth]{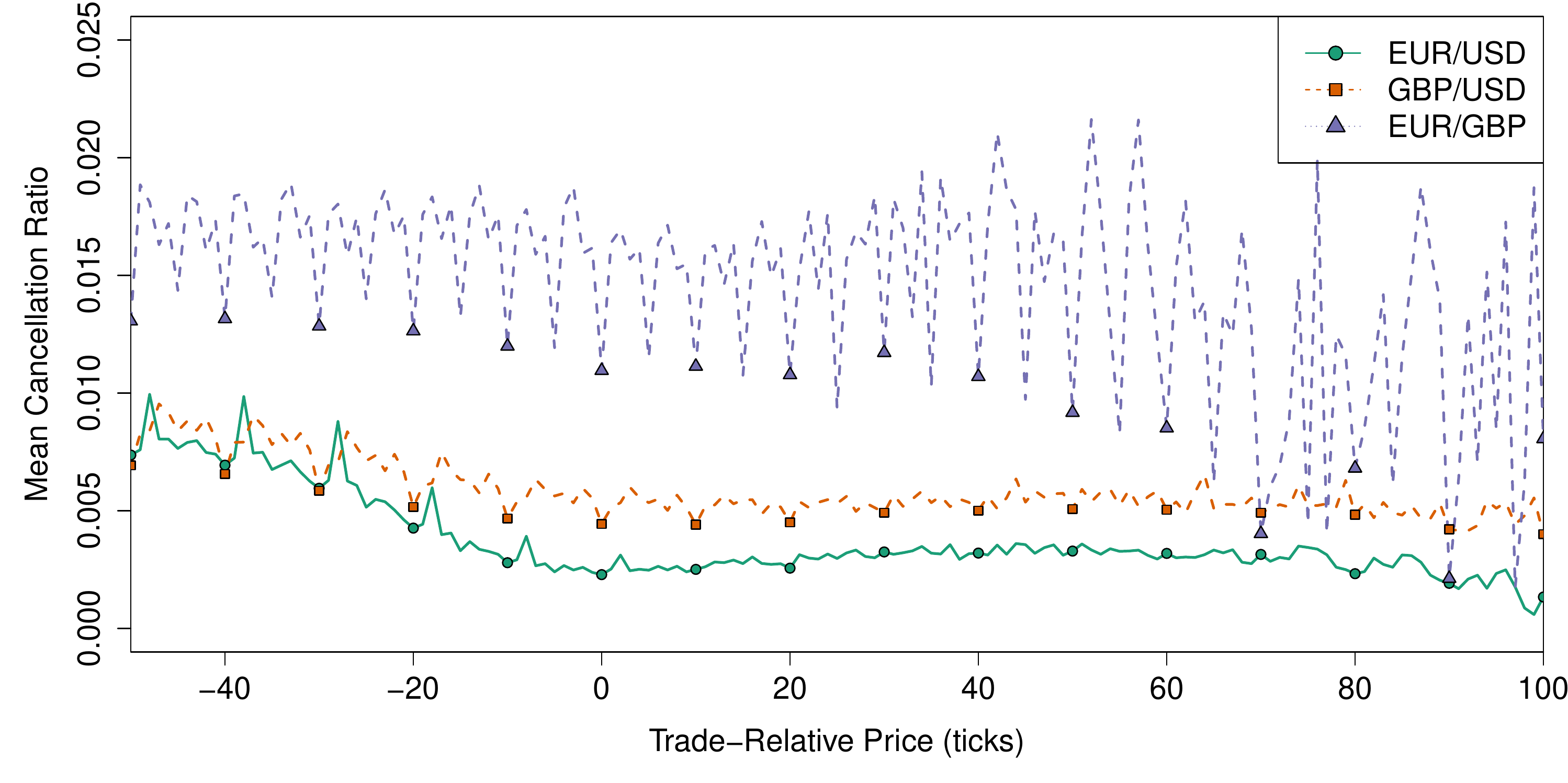}
\caption{Mean cancellation ratio (i.e., total size of cancelled active orders at a given relative price divided by the corresponding mean depth) at given (top) quote-relative and (bottom) trade-relative prices for (solid green curves with circles) EUR/USD, (dashed orange curves with squares) GBP/USD, and (dotted purple curves with triangles) EUR/GBP on 4 May 2010.}
\label{fig:sfxCPOday}
\end{figure}

We obtain additional insights into round-number effects by calculating the \emph{mean cancellation ratio,} which we measure by rescaling the total size of cancelled active orders at a given relative price by the corresponding mean depth (see Figure \ref{fig:sfxCPOday}). The mean cancellation ratio is a useful quantity for helping to understand order cancellations, because simply calculating the total size of active order cancellations at a given relative price (as in Figure \ref{fig:sfxCXdistday}) does not take into account that the mean depth, and therefore the mean total size of active orders that \emph{could} be cancelled, varies substantially across relative prices (see Figure \ref{fig:sfxDPdistday}).

In quote-relative coordinates, the mean cancellation ratios vary considerably with relative price, with no discernible trend or pattern. However, this is unsurprising because institutions in a QCLOB are unable to calculate quote-relative prices and therefore cannot use such information when deciding whether to cancel an order. In trade-relative coordinates, by contrast, two interesting results emerge. First, each of the three currency pairs' mean cancellation ratios exhibit a strong round-number periodicity: the mean lifetime of an active order at a trade-relative price that is an integer multiple of 5 is longer than that of an active order at a neighbouring trade-relative price. Second, aside from this round-number effect, the mean cancellation ratios for EUR/USD and GBP/USD are approximately constant for negative trade-relative prices and decrease for positive trade-relative prices. At all trade-relative prices, the cancellation ratio for EUR/GBP is higher than it is for the other two currency pairs. However, the round-number effect is particularly strong for EUR/USD, so it is difficult to discern the variation in mean cancellation ratio across trade-relative prices.

\subsection{Comparisons Across Trading Days}\label{subsec:differentdays}

\begin{figure}[!htbp]
\centering
\includegraphics[width=0.8\textwidth]{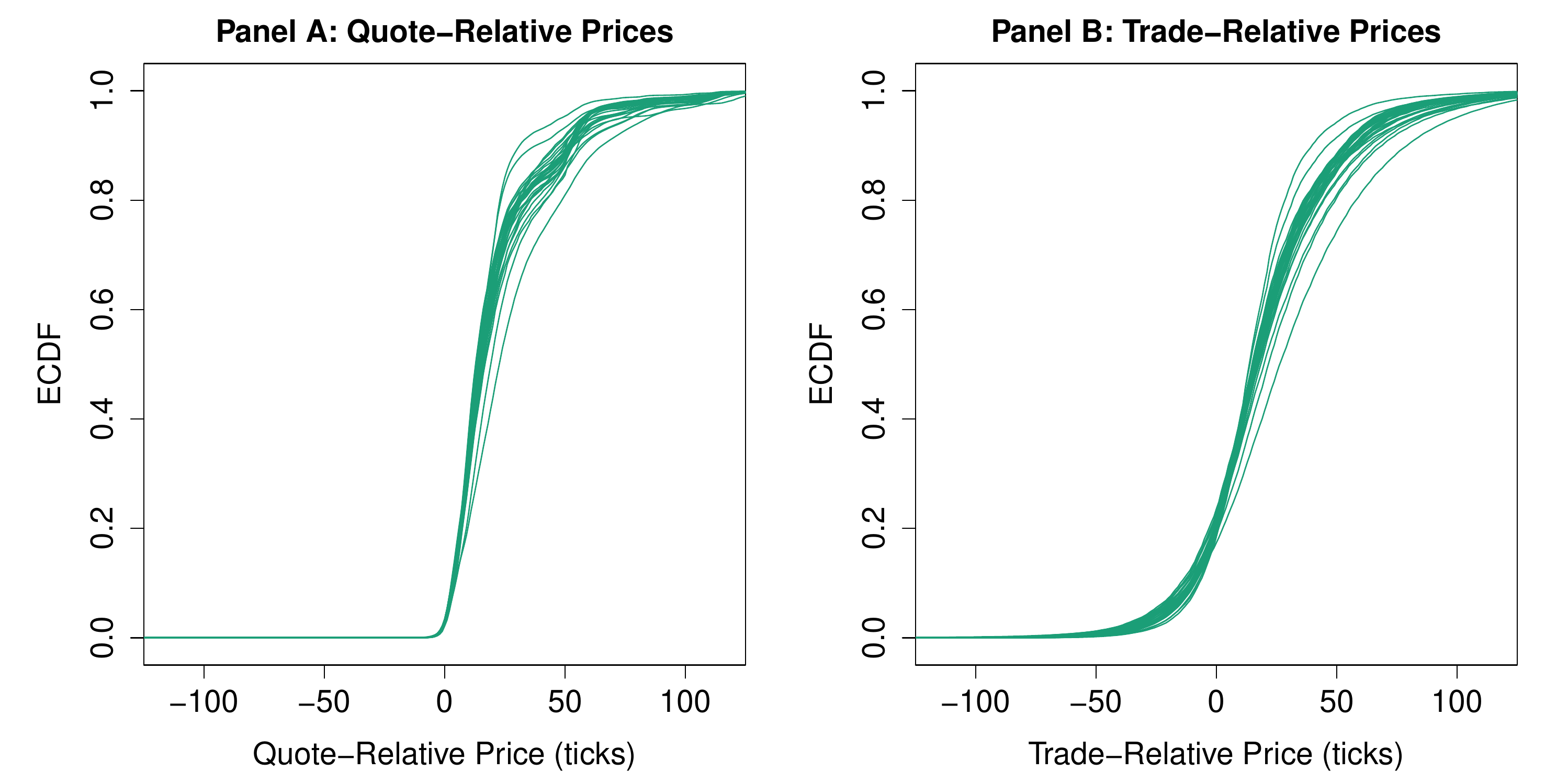}
\caption{ECDFs for EUR/USD limit order arrivals at given (left panel) quote-relative and (right panel) trade-relative prices. Each curve indicates the ECDF on a single day.}
\label{fig:sfxDailyECDFs}
\end{figure}

We now investigate how the distributions of order flow and LOB state vary across trading days. In Figure \ref{fig:sfxDailyECDFs}, we show the ECDFs of limit order arrivals for EUR/USD. Each curve indicates the given distribution for a single trading day. The results for cancellations and normalized mean depths, and the corresponding results for the other currency pairs, are all qualitatively similar. In each case, the ECDFs suggest that there are substantial differences across different days. On some days, the majority of order arrivals and cancellations occur over a narrow range of small relative prices; on other days, the range of relative prices over which such activity occurs is wider, which indicates that a larger fraction of activity occurs deeper into the global LOB.

\begin{figure}[!htbp]
\centering
\includegraphics[width=0.8\textwidth]{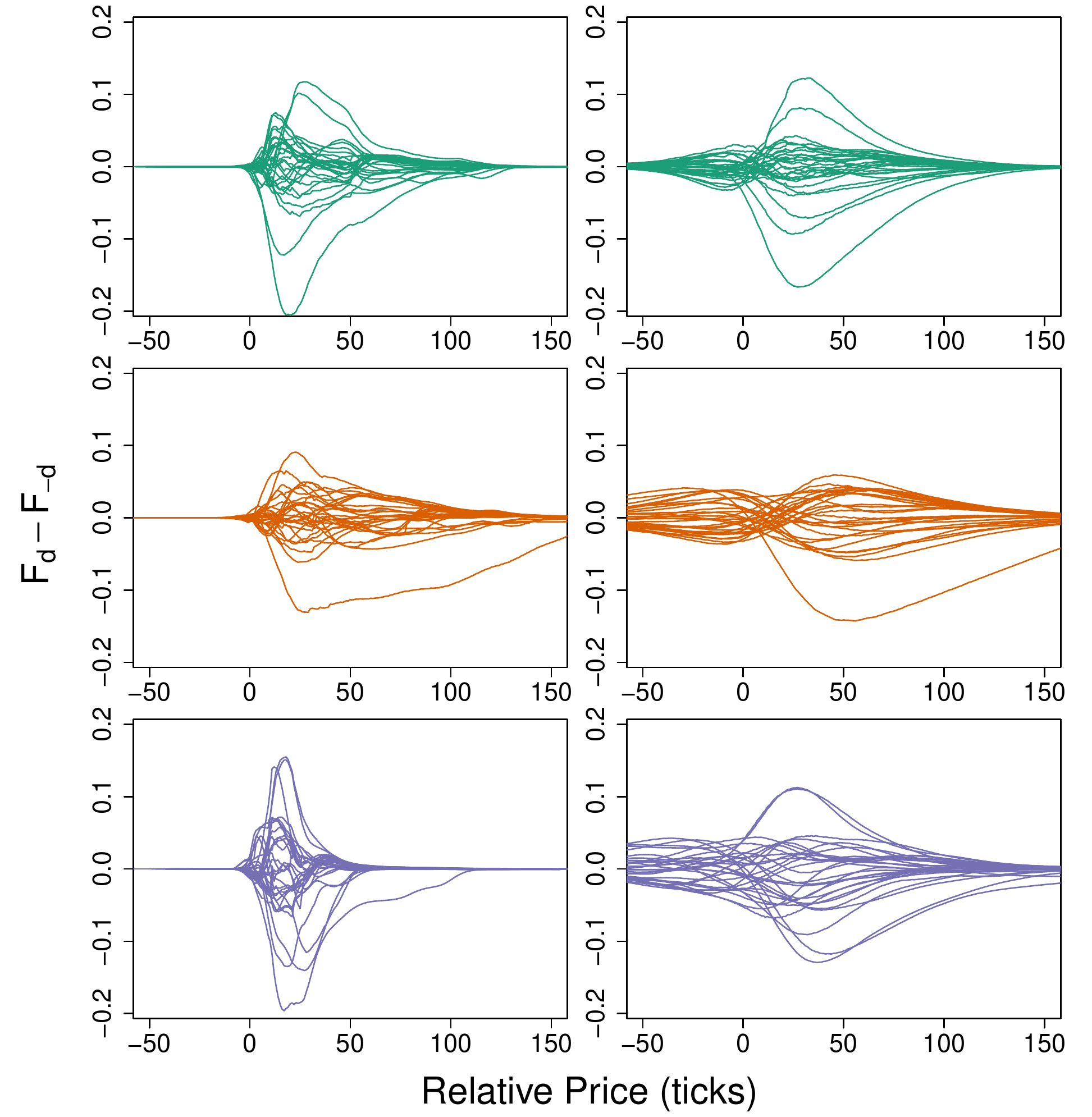}
\caption{Distances between ECDF $F_d$ of limit order arrivals on a given day $d$ and ECDF $F_{-d}$ of limit order arrivals on all other days at given (left panel) quote-relative and (right panel) trade-relative prices for (green curves) EUR/USD, (orange curves) GBP/USD, and (purple curves) EUR/GBP. Each curve indicates the distances for a single day $d$.}
\label{fig:sfxDailyECDFDistances}
\end{figure}

To help quantify the differences between these daily distributions, we also calculate the distance between a given day's ECDF and the corresponding ECDF for the aggregate data from all other 29 days in our sample. For example, when studying EUR/USD limit order arrivals on 4 May 2010, we first calculate the ECDF using the data for just this day (as in Figure \ref{fig:sfxDailyECDFs}) and then calculate the ECDF for EUR/USD limit order arrivals on all other days in our sample. We write $F_d(p)$ to denote the ECDF for the data on day $d$, and we write $F_{-d}(p)$ to denote the ECDF for the data on all days except day $d$. We then calculate the difference $F_d(p)-F_{-d}(p)$. In Figure \ref{fig:sfxDailyECDFDistances}, we show the resulting plots for limit order arrivals. The results for cancellations and normalized mean depths are qualitatively similar. As also illustrated in Figure \ref{fig:sfxDailyECDFs}, the distributions on individual trading days often differ substantially from the aggregate distributions from the other trading days.

To investigate the extent to which differences in the first two moments account for the observed differences between the daily distributions, we rescale each day's data according to its sample mean and standard deviation. When calculating these sample moments, we use a trimmed sample mean and trimmed sample standard deviation to exclude all order arrivals and cancellations that occur with a relative price of more than $1000$ ticks.\footnote{For a detailed discussion of trimmed sample moments, see \cite{Huber:2009robust}.} This trimming removes a very small number of orders with extremely large relative prices. For example, all EUR/USD trades in the data occur in the price interval $\$1.10$--$\$1.40$, but some sell limit orders arrive with a price of more than $\$500.00$. Such orders do not seem to represent a serious intention to trade. For each of the three currency pairs, for both buy and sell orders, and in both quote-relative and trade-relative coordinates, trimming the data in this way removes less than $0.05\%$ of the total order flow. We also obtain qualitatively similar results if we instead trim all orders whose relative prices are within the the top $1$ percentile of the respective distributions.

\begin{figure}[!htbp]
\centering
\includegraphics[width=0.8\textwidth]{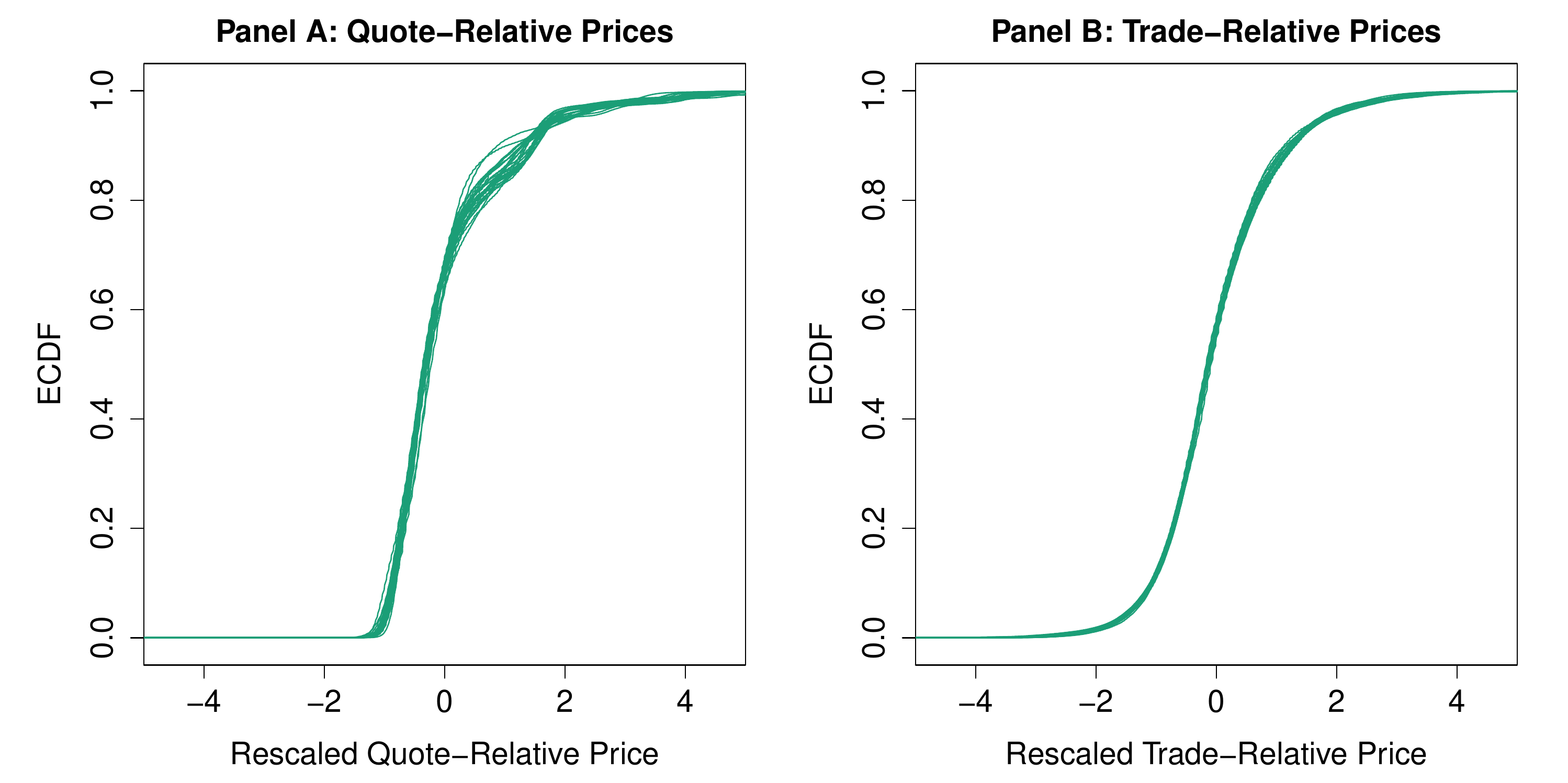}
\caption{ECDFs for EUR/USD limit order arrivals at given (left panel) quote-relative and (right panel) trade-relative prices after rescaling each day's data by subtracting its sample mean and dividing by its sample standard deviation. Each curve indicates the ECDF on a single day.}
\label{fig:sfxDailyECDFsRescaled}
\end{figure}

In Figure \ref{fig:sfxDailyECDFsRescaled}, we show the ECDFs of EUR/USD limit order arrivals after rescaling the data to account for the daily differences in its first two moments. The results for the other currency pairs are qualitatively similar. In quote-relative coordinates, the rescaling causes a reasonably strong collapse for limit order arrivals and cancellations, but daily differences in the distributions' upper tails prevents a stronger collapse in this region. In trade-relative coordinates, the rescaling causes a strong collapse onto what appears to be a single, universal curve over the whole domain. In both quote-relative and trade-relative coordinates, the collapse for the distributions of normalized mean depths is slightly weaker than for the order-flow distributions due to a handful of orders with extremely large relative prices that remain active for long periods on some days.\footnote{To verify that such extreme-priced orders are indeed the primary reason for the weaker collapse of these distributions, we repeated our calculations after excluding all active orders with a relative price of more than 5 standard deviations from the mean. We found that the resulting curve collapse was similar to that for limit order arrivals and cancellations.}

To investigate the strength of this curve collapse, we calculate the distance between a given day's ECDF and the corresponding ECDF for the aggregate data from all other 29 days in our sample (as in Figure \ref{fig:sfxDailyECDFDistances}) after performing the rescaling to account for daily differences in the first two moments. Specifically, for a given day $d$, we first rescale the data from each of the other 29 days by subtracting each day's sample mean and dividing by its sample standard deviation. We then aggregate the rescaled data from these 29 days, multiply the result by the sample standard deviation on day $d$, and add the sample mean on day $d$. Finally, we calculate this rescaled, aggregated data set's ECDF, which we label $\hat{F}_{-d}$, and we then calculate its distance from $F_{d}$. We perform our calculations in this way to ensure that the domain of our distance measurements matches that of the data from day $d$. This enables us to perform direct comparisons to our results for the non-rescaled data.

\begin{figure}[!htbp]
\centering
\includegraphics[width=0.8\textwidth]{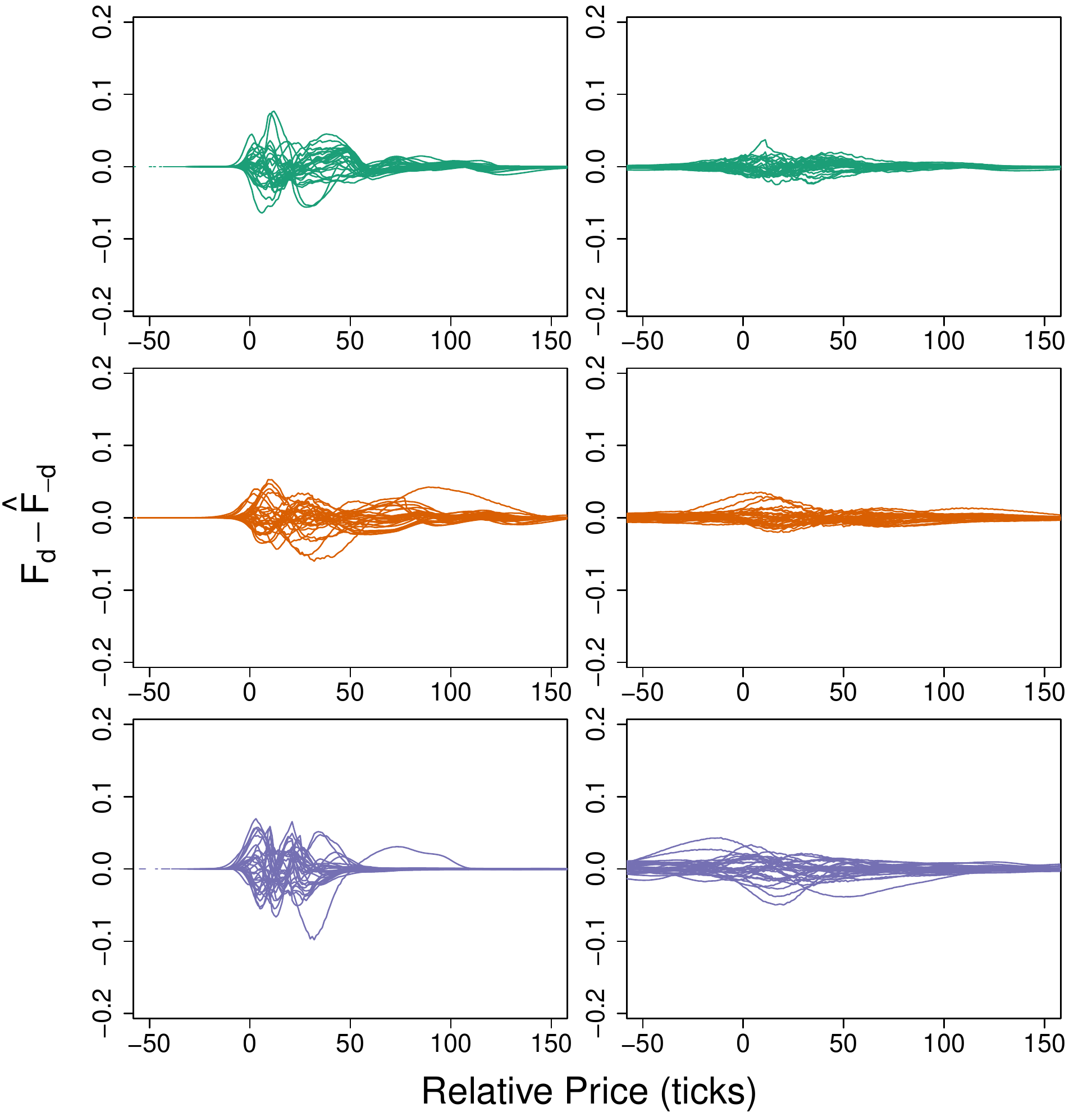}
\caption{Distances between ECDF $F_d$ of limit order arrivals on a given day $d$ and rescaled ECDF $\hat{F}_{-d}$ of limit order arrivals on all other days, at given (left panel) quote-relative and (right panel) trade-relative prices for (green curves) EUR/USD, (orange curves) GBP/USD, and (purple curves) EUR/GBP. Each curve indicates the distances for a single day $d$.}
\label{fig:sfxDailyRescaledECDFs}
\end{figure}

In Figure \ref{fig:sfxDailyRescaledECDFs}, we show the distances $F_d-\hat{F}_{-d}$ for limit order arrivals; the results for cancellations and normalized mean depths are qualitatively similar. In quote-relative coordinates, rescaling the data to account for daily differences in the first two moments produces a considerable reduction in distances between the daily ECDFs. This reduction is particularly strong for the days whose distributions are furthest from the aggregate distribution across the other days (see Figure \ref{fig:sfxDailyECDFDistances}). In trade-relative coordinates, the rescaling causes very strong curve collapse across the entire domain and on all days.

To quantify the strength of this curve collapse, we compute the mean ratio $\overline{C}$ of the Cram\'{e}r--von Mises (CvM) distances \cite{Cramer:1928composition,Huber:2012goodness} between the distributions before and after applying the rescaling (see Table \ref{tab:sfxoosperformance}).\footnote{We also find qualitatively similar results when using the Kolmogorov--Smirnov (KS) distance \cite{Smirnov:1939estimation,Wasserman:2004statistics}. There are many other possible distance measures \cite{Deza:2006dictionary} that we could use; we choose the CvM and KS distances because they are widely used, easy to interpret, and fast to compute.} We give a detailed discussion of our methodology in Appendix \ref{app:curvecollapse}. 

\begin{table}[!htbp]
\begin{center}
\begin{tabular}{|l|l|l|l|l|}
\hline
Coordinates & Order Flow & EUR/USD & GBP/USD & EUR/GBP \\
\hline
\multirow{3}{*}{Quote relative} & Limit orders & $4.36$ & $4.10$ & $5.11$ \\ 
& Cancellations & $3.92$ & $3.83$ & $4.92$ \\ 
& Mean depths & $1.45$ & $2.03$ & $2.83$ \\ 
\hline
\multirow{3}{*}{Trade relative} & Limit orders & $20.73$ & $25.13$ & $21.78$ \\ 
& Cancellations & $20.20$ & $24.37$ & $21.65$ \\ 
& Mean depths & $3.04$ & $10.06$ & $11.07$ \\ 
\hline
\end{tabular}
\caption{Mean CvM ratios $\overline{C}$ (see the description in the main text and in Appendix \ref{app:curvecollapse}) for limit order arrivals, cancellations, and mean depths. Values larger than 1 indicate that rescaling each day's data to account for differences in its first two moments reduces the mean distance between the daily distributions. Larger values correspond to stronger curve collapse.}
\label{tab:sfxoosperformance}
\end{center}
\end{table}

In quote-relative coordinates, the reductions in CvM distance for limit orders and cancellations range from a factor of about 4 to a factor of about 5. This indicates a moderately strong curve collapse. The corresponding reductions for normalized mean depths are weaker because of a small number of extreme-priced orders that remain active for long periods of time and thereby prevent stronger collapse in the upper tails of these distributions. In trade-relative coordinates, the reductions in CvM distance for limit order arrivals and cancellations range from a factor of about 20 to a factor of more than 25. This indicates very strong curve collapse. Again, the corresponding reductions for normalized mean depths are weaker (particularly for EUR/USD), but they still indicate a moderate curve collapse for EUR/USD and a strong curve collapse for GBP/USD and EUR/GBP.

\subsection{Models of Order Flow and LOB State}

In recent years, many authors have studied simple models of order flow and LOB state to help understand the complex dynamics that occur in financial markets (see \cite{Gould:2013limit}). When constructing such models, it is often desirable to incorporate simple, statistical descriptions of order flow and LOB state that capture the salient features of real market activity. In this section, we use our results from the previous sections to motivate two approaches to this problem in a QCLOB.

The first approach that we consider is a parametric approach. In their study of order flow on the LSE, \cite{Mike:2008empirical} used a generalized $t$ distribution to model the distributions of quote-relative prices of arriving orders. For order flow and LOB state on Hotspot FX, we find that this distribution provides a moderate fit in quote-relative coordinates and a strong fit in trade-relative coordinates. Several other parametric distributions with more than four parameters (most notably, the five-parameter logistic distribution \cite{Gottschalk:2005five}) also fit the data well, but the inclusion of additional parameters increases the computational complexity of the required optimization, and could also lead to over-fitting. We therefore restrict our attention to the generalized $t$ distribution.

\begin{figure}[!htbp]
\centering
\includegraphics[width=0.8\textwidth]{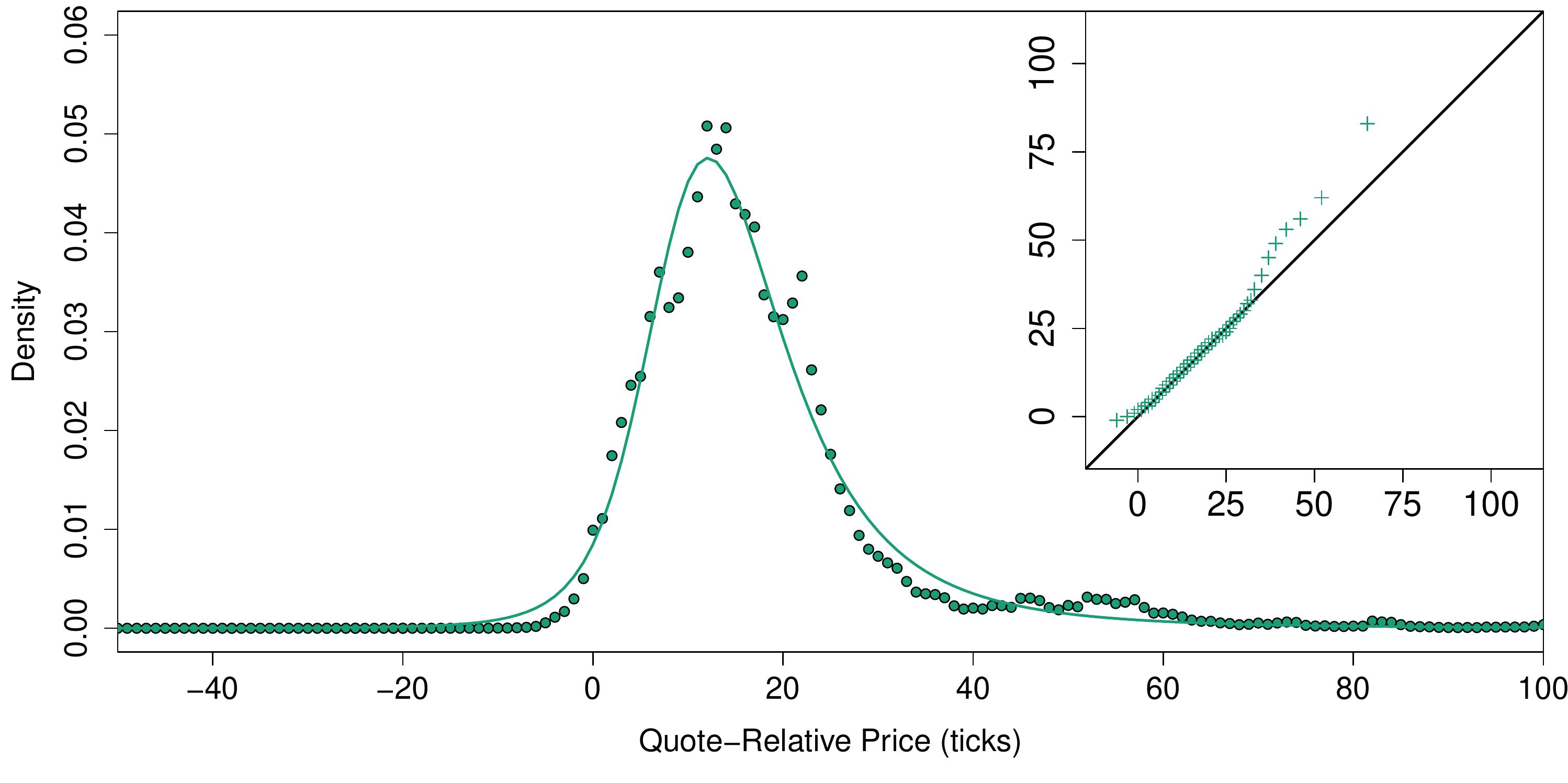}
\includegraphics[width=0.8\textwidth]{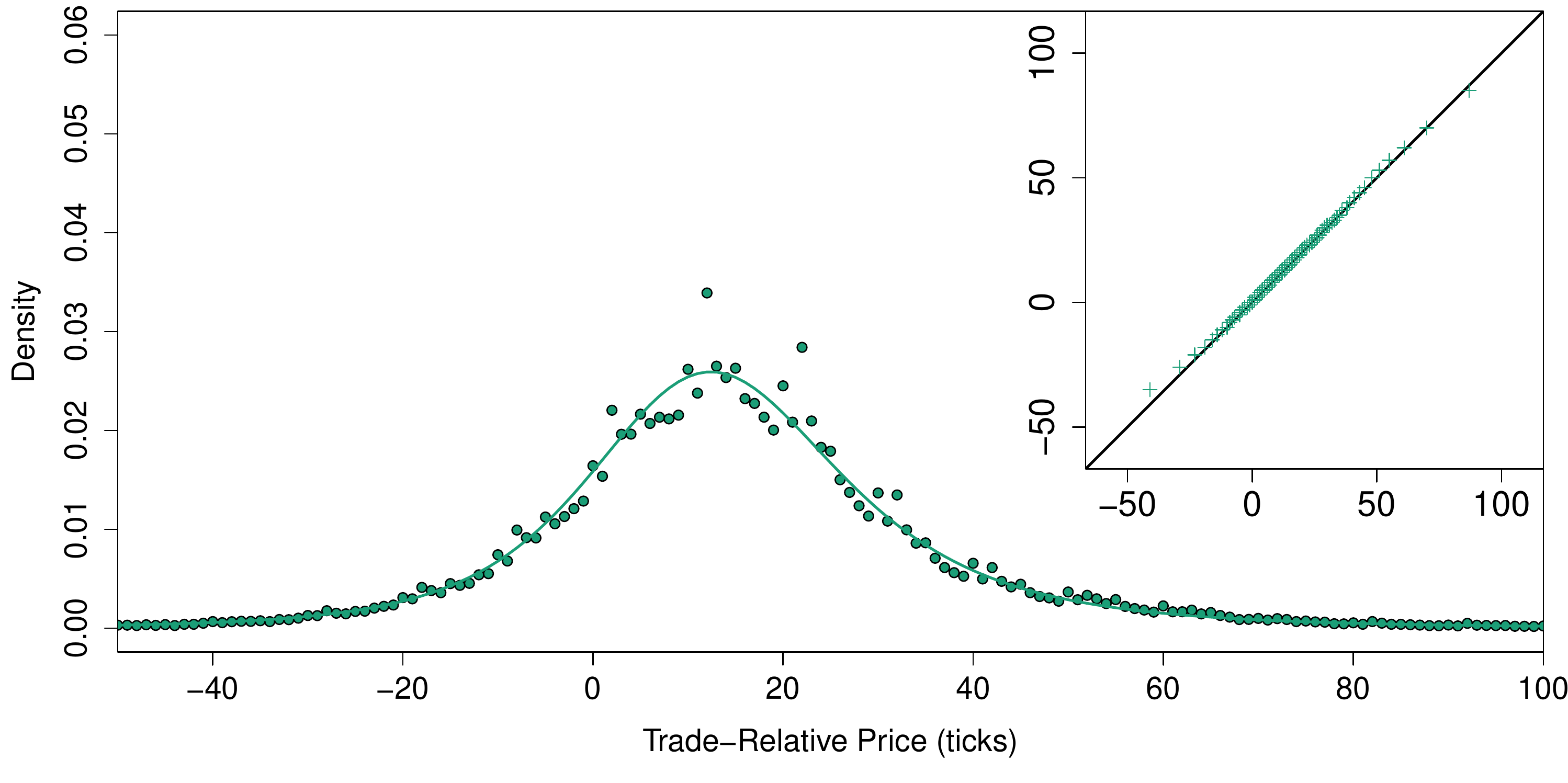}
\caption{Fits of the generalized $t$ distribution to the distribution of limit order arrivals for EUR/USD on 4 May 2010 in (top) quote-relative and (bottom) trade-relative coordinates. The main plots show (green circles) the empirical density functions and (green curves) our fits of the generalized $t$ distribution. In the inset, we show quantile-quantile (Q-Q) plots of (vertical axis) the ECDFs versus (horizontal axis) our fits of the generalized $t$ distribution. The points indicate the $1^{\text{st}}, 2^{\text{nd}}, \ldots, 99^{\text{th}}$ percentiles of the distributions. The solid black lines indicate the diagonal. The results for the other currency pairs are qualitatively similar.}
\label{fig:sfxLOfit}
\end{figure}

In Figure \ref{fig:sfxLOfit}, we show our fit of the generalized $t$ distribution to the quote-relative and trade-relative distributions of limit order arrivals for EUR/USD on 4 May 2010. We describe our method of fitting the distribution in Appendix \ref{app:fittingt}. The results for the other currency pairs and other dates are qualitatively similar. Although the distribution fails to capture some of the features of the order flow that we observe on Hotspot FX (such as the tendency for orders to arrive more frequently at round-number relative prices), the fits perform reasonably well. In quote-relative coordinates, the fits match the approximate shape of the empirical density in the middle of the domain, but they fail to capture the strong kurtosis of the data, and they therefore do not perform very well in the upper and lower tails. In trade-relative coordinates, the fits perform well over the whole domain.

\begin{figure}[!htbp]
\centering
\includegraphics[width=0.8\textwidth]{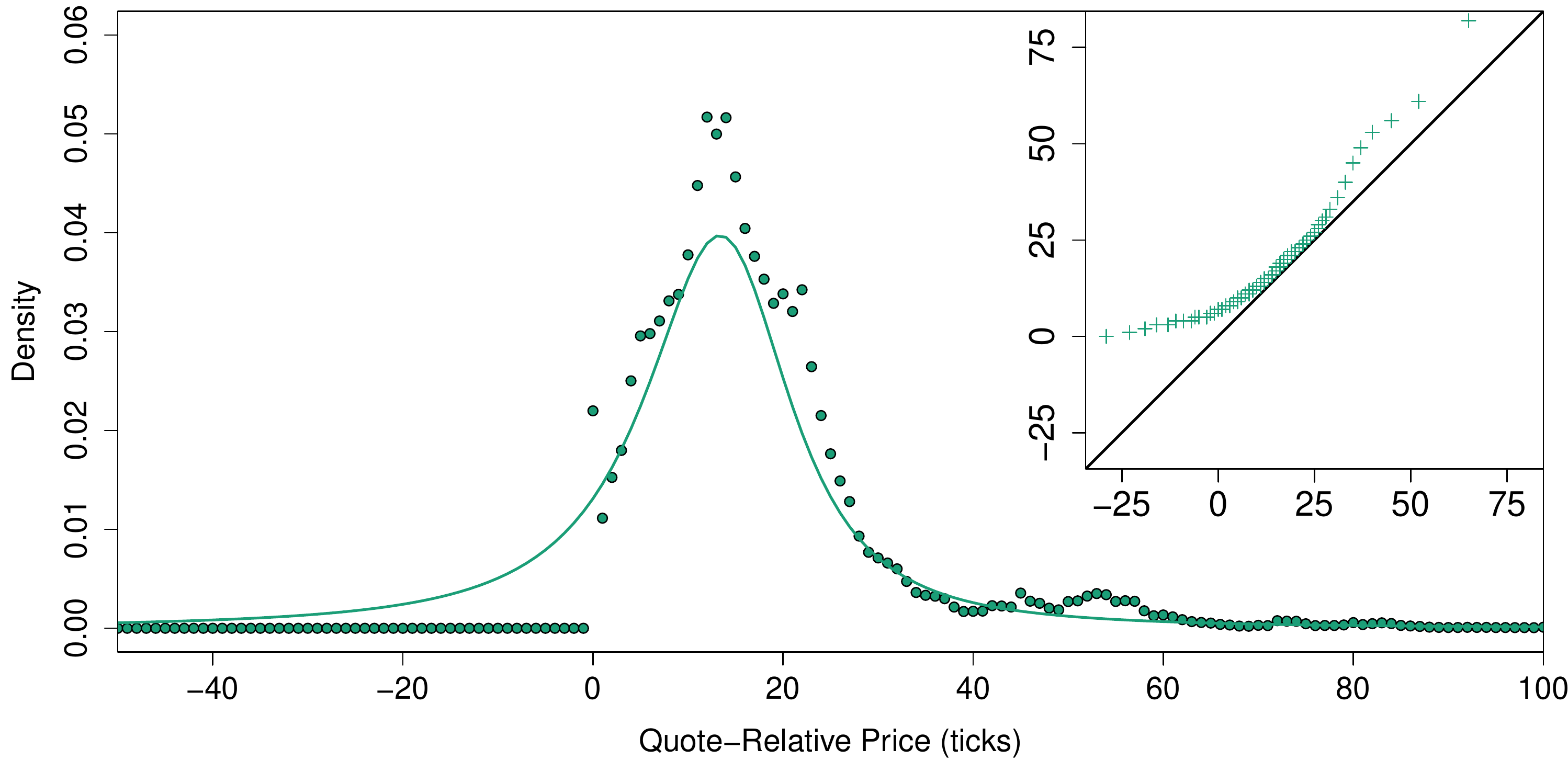}
\includegraphics[width=0.8\textwidth]{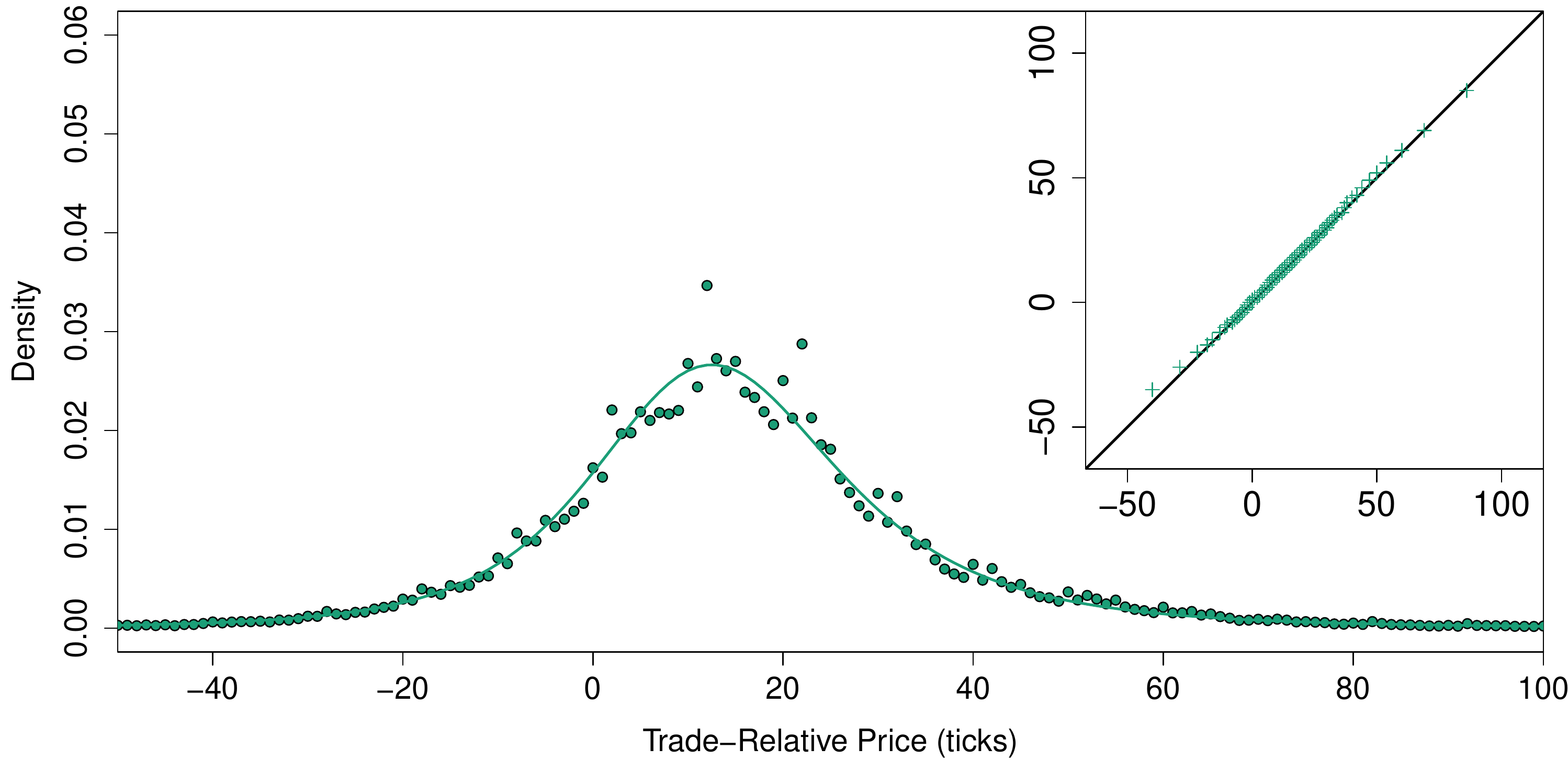}
\caption{Fits of the generalized $t$ distribution to the distribution of cancellations for EUR/USD on 4 May 2010 in (top) quote-relative and (bottom) trade-relative coordinates. The main plots show (green circles) the empirical density functions and (green curves) our fits of the generalized $t$ distribution. In the inset, we show Q-Q plots of (vertical axis) the ECDFs versus (horizontal axis) our fits of the generalized $t$ distribution. The points indicate the $1^{\text{st}}, 2^{\text{nd}}, \ldots, 99^{\text{th}}$ percentiles of the distributions. The solid black lines indicate the diagonal. The results for the other currency pairs are qualitatively similar.}
\label{fig:sfxCXfit}
\end{figure}

In trade-relative coordinates, we again find that a generalized $t$ distribution provides a good fit to the distribution of active order cancellations (see Figure \ref{fig:sfxCXfit} for EUR/USD on 4 May 2010; the results for the other currency pairs and other dates are all qualitatively similar). In quote-relative coordinates, the local maximum in active order cancellations at a quote-relative price of 0 hinders this approach because the shape of the generalized $t$ distribution does not capture this feature of the data. Therefore, the fits for quote-relative cancellations are outperformed by the fits for quote-relative limit order arrivals (see Figure \ref{fig:sfxLOfit}). The results for the normalized mean depths are qualitatively similar to those for cancellations, so we omit these plots.

The results in Figure \ref{fig:sfxDailyRescaledECDFs} and Table \ref{tab:sfxoosperformance} also motivate an alternative, semi-parametric approach to modelling the distributions of order flow and LOB state. For a single trading day $d$, let $\mu_d$ and $\sigma_d$ denote, respectively, the mean and standard deviation of a specified property (e.g., EUR/USD limit order arrivals in trade-relative coordinates). Given data from a set $D$ of trading days, we rescale the data on each day $d$ by subtracting $\mu_d$ then dividing by $\sigma_d$, and we then aggregate the rescaled data for all days into a single data set. To obtain the model for the distribution on another trading day $d' \notin D$, we multiply each entry in the aggregated data set by $\sigma_{d'}$ then add $\mu_{d'}$.

\begin{figure}[!htbp]
\centering
\includegraphics[width=0.8\textwidth]{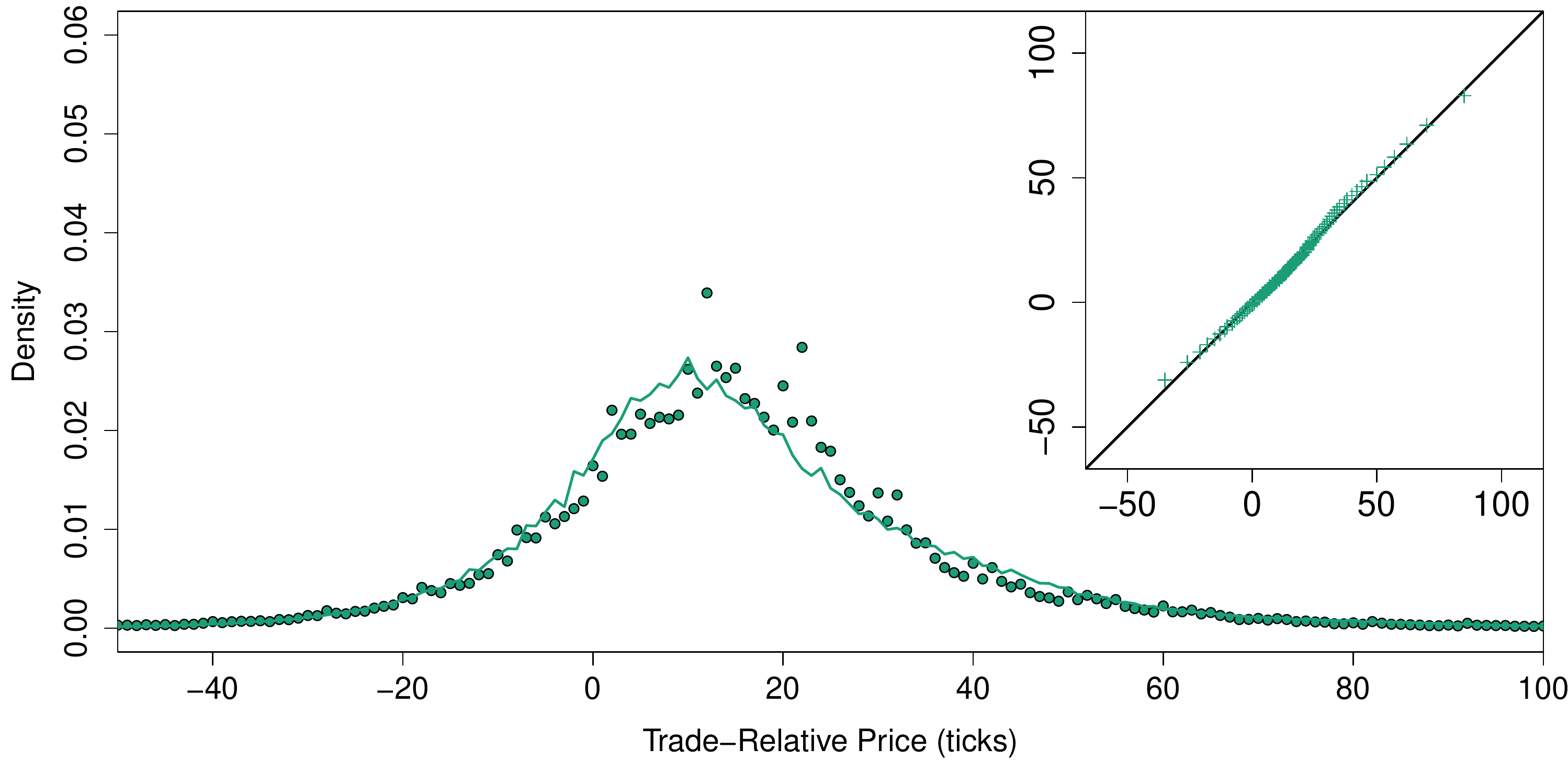}
\caption{Semi-parametric fit of the distribution of limit order arrivals for EUR/USD on 4 May 2010 in trade-relative coordinates. The main plot shows (green circles) the empirical density function for limit orders and (solid green curve) the corresponding semi-parametric fit obtained by rescaling and aggregating the data from all other days in our sample then inverting the rescaling according to the mean and standard deviation on 4 May 2010. In the inset, we show a Q-Q plot of (vertical axis) the ECDFs versus (horizontal axis) our semi-parametric fits of the distribution. The points indicate the $1^{\text{st}}, 2^{\text{nd}}, \ldots, 99^{\text{th}}$ percentiles of the distributions. The solid black line indicates the diagonal. The results for cancellations, for the other currency pairs, and for the other days in our sample are all qualitatively similar.}
\label{fig:sfxLONonparametric}
\end{figure}

In Figure \ref{fig:sfxLONonparametric}, we show the result of applying this semi-parametric approach to model the trade-relative distribution of limit order arrivals for EUR/USD on 4 May 2010. The results for cancellations, for the other currency pairs, and for the other days in our sample are all qualitatively similar. As illustrated by the Q-Q plot, the fit performs well over the whole domain. The corresponding fits for normalized mean depths and for the distributions in quote-relative coordinates perform slightly less well because of a small number of extreme-priced orders in the upper tail (see Figure \ref{fig:sfxDPdistday}), but given that such activity corresponds to limit orders with very low fill probabilities, we do not regard a close fit in this region to be as important as it is for the main body of the distribution, where the fits are strong.

In all cases, the performance of our semi-parametric method is similar to that of fitting the generalized $t$ distribution directly to the data (see Figure \ref{fig:sfxLOfit}). However, our semi-parametric approach offers a considerable computational advantage: after computing the aggregated data set -- which, given a historical database of trading days, can be performed offline and in advance of fitting a single trading day -- performing the semi-parametric fit requires only a multiplication and an addition. By contrast, fitting the generalized $t$ distribution requires numerical optimization of a nonlinear objective function (see Appendix \ref{app:fittingt}), which is much slower to perform.

In some applications, the simplicity of employing a well-known parametric distribution may outweigh the possible gains of our semi-parametric approach. In others, the reduction in computational overhead offered by our semi-parametric approach may outweigh the benefits of using a parametric distribution. Therefore, we anticipate that both of these approaches will be useful in different contexts.

\section{Discussion}\label{sec:sfxdiscussion}

In this section, we address several interesting points raised by our results, and we compare our findings for QCLOBs to those reported by empirical studies of centralized LOBs in order to highlight some important differences between these two market organizations.

One important difference between QCLOBs and centralized LOBs is the shape of the distributions of order flow. Several empirical studies of centralized LOBs have reported that the maximum limit order arrival rate occurs at a quote-relative price of 0 \cite{Biais:1995empirical,Bouchaud:2002statistical,Gu:2008empiricalregularities,Hollifield:2004empirical,Mike:2008empirical,Potters:2003more}, whereas the maximum limit order arrival rate on Hotspot FX occurs at a strictly positive quote-relative price (see Figure \ref{fig:sfxLOdistday}). We propose the following explanation for this observation. In a QCLOB, each institution $\theta_i$ sees the values of $b_i(t)$ and $a_i(t)$, but cannot see the values of $b(t)$ and $a(t)$. By definition, $b_i(t)\leq b(t)$ and $a_i(t) \geq a(t)$, so if each institution bases its trading decisions on $b_i(t)$ and $a_i(t)$, and if $b_i(t)$ and $a_i(t)$ both typically reside at strictly positive quote-relative prices, then the maximum arrival rate of the aggregate limit order flow generated by all institutions will occur at a strictly positive quote-relative price.

Similarly, several empirical studies of centralized LOBs have reported that cancellations occur most often among active orders at $b(t)$ and $a(t)$, and less often among active orders deeper into the LOB \cite{Cont:2010stochastic,Potters:2003more}. Several authors have conjectured that the high number of cancellations at these prices indicate that many institutions compete for priority at the best quotes, and that the lower cancellation rates among other orders indicate that their owners aim to profit from large price movements on longer time horizons \cite{Challet:2001analyzing,Potters:2003more,Zovko:2002power}. We also observe a local maximum in the distribution of cancellations at a quote-relative price of 0 (see Figure \ref{fig:sfxCXdistday}), but we find the distribution's global maximum to be strictly positive. After rescaling to account for differences in the mean depths, we find that the quote-relative cancellation ratios vary considerably, with no clear trend (see Figure \ref{fig:sfxCPOday}). In trade-relative coordinates, we find that the cancellation distributions closely resemble those of limit order arrivals, with a slightly lower cancellation ratio among orders with larger trade-relative prices.

Many centralized LOBs have been reported to exhibit a ``hump'' shape that first increases and then subsequently decreases away from the best quotes \cite{Bouchaud:2002statistical,Gu:2008empiricalshape,Hollifield:2004empirical,Potters:2003more}. \cite{Rosu:2009dynamic} conjectured that such a hump represents a trade-off between an optimism that limit orders placed far from the spread may eventually result in a significant profit and a pessimism that such orders may never match. We also observe a hump shape in the mean state of $\mathcal{L}(t)$ in both quote-relative and trade-relative coordinates (see Figure \ref{fig:sfxDPdistday}). For the LOBs examined in other empirical studies, however, market orders accounted for about $10\%$--$30\%$ of the total arriving order flow, and they therefore played an important role in maintaining the hump shape of $\mathcal{L}(t)$ \cite{Challet:2001analyzing,Gereben:2010brief,Hasbrouck:2002limit,Lo:2010order,Potters:2003more}. On Hotspot FX, market orders constitute less than $0.05\%$ of the total arriving order flow (see Table \ref{tab:sfxallagg}). Therefore, the hump shapes that we observe are primarily a consequence of similar shapes in the distributions of limit order arrivals and cancellations.

Why do institutions submit so many limit orders, given that so few result in trades? We propose two possible explanations. First, institutions may place limit orders on several different trading platforms simultaneously to increase their chance of receiving a matching. If one such order matches, an institution can simply cancel the duplicates on other platforms. \cite{Cont:2014optimal} recently noted that this strategy, which they called ``overbooking'', becomes more prominent as the number of venues for a given asset increases. In some markets, overbooking exposes an institution to the risk of receiving near-simultaneous matchings on multiple platforms, but several FX spot trading platforms (including Hotspot FX) allow liquidity providers to apply a ``last look'' feature to their limit orders. This feature enables liquidity providers to reject an incoming market order after it arrives.\footnote{For a detailed discussion of last look, see \cite{Cartea:2015foreign}.} Even though the total volumes of trade on Hotspot FX are very large, they constitute only a small fraction of the total volumes across all electronic trading platforms in the FX spot market (see Section \ref{subsec:hfx}). Together, the availability of alternative trading opportunities on other platforms and the protection offered by last look against unintended matches make overbooking extremely attractive, and could therefore result in a large volume of cancellations from institutions that adopt this strategy. Second, many high-frequency and algorithmic trading techniques involve the submission and cancellation of large numbers of limit orders \cite{Biais:2011equilibrium, Chaboud:2011rise, Hendershott:2011algorithmic, Kirilenko:2011flash}. The recent surge in popularity of trading strategies that utilize such techniques could account for a high percentage of the cancellations that we observe.

Another important difference between QCLOBs and centralized LOBs is the possibility for the appearance of market configurations that would not be possible in a centralized LOB. On Hotspot FX, we observe a negative global spread reasonably often for EUR/USD and GBP/USD (see Table \ref{tab:sfxspread}). This observation motivates another question: What fraction of the global liquidity in $\mathcal{L}(t)$ can an institution $\theta_i$ typically access in its local LOB $\mathcal{L}_i(t)$? Although the Hotspot FX data does not provide a way to reconstruct local LOBs for specific institutions, several of our results and observations provide insight into this question.

First, as noted above, we observe periods during which the global bid--ask spread for a given currency pair is negative for several seconds. This suggests that among the institutions that place limit orders close to the best quotes, there exist some pairs of institutions, $\theta_j$ and $\theta_k$, such that no other institution $\theta_i$ is able to access the limit orders posted by both $\theta_j$ and $\theta_k$. Otherwise, $\theta_i$ would submit a pair of buy and sell market orders to capitalize on the arbitrage opportunity, and would thereby widen the spread to a non-negative size. In Section~\ref{subsec:qclobs}, we exhibited a toy QCLOB (suggestive of a core--periphery structure) in which negative global bid--ask spreads are caused by institutions that have relatively poor CCLs. If an institution with only one trading partner posts a limit order that exceeds its bilateral CCL, at least part of that order will be unseen by any other institution at all, and could in principle cause an arbitrarily large negative bid--ask spread.

Second, we also observe surprising results when studying selective liquidity-taking on Hotspot FX (see Figure \ref{fig:sfxLOMOfraction}). Institutions appear to condition their market order sizes on the depth available when this depth is small, but they appear not to do so when this depth is large. One possible explanation is that when an institution $\theta_i$ seeks to submit a buy (respectively, sell) market order, if the total depth of active orders at $a_i(t)$ (respectively, $b_i(t)$) is larger than $\theta_i$'s desired market order size, then it may no longer be necessary for the institution to condition its order size according to the available liquidity. However, in a similar study of selective liquidity-taking on the LSE (which operates as a centralized LOB), \cite{Farmer:2004what} reported the approximately linear relationship that we observe for smaller queue lengths to persist across the whole domain (i.e., even when the queue length is very large). An alternative explanation is that the effect that we observe is a consequence of the CCLs in a QCLOB, and specifically that some institutions are only able to access a relatively small fraction of the active orders at a given price in the global LOB. When the depth at $a_i(t)$ (respectively, $b_i(t)$) is small, it is likely to consist of a single active order. In this scenario, the linear relationship that we observe for small queue lengths could be caused by $\theta_i$ conditioning its market order size according to the size of this single active order. When the depth at $a_i(t)$ (respectively, $b_i(t)$) is larger, however, it is more likely to consist of several different active orders, each with a different owner. Because the Hotspot FX data describes the global LOB $\mathcal{L}(t)$, we are able to see all such orders at the given price. However, a given institution $\theta_i$ that trades on the platform can only see the subset of these orders that are owned by other institutions with whom it has sufficient CCLs. Therefore, $\theta_i$ may only see a small subset of the liquidity that is available globally, and it may therefore condition its market order size according to the depth that it sees.

Third, the ratio of the mean total size of market orders on a single day to the mean total size of active orders (which is often used as a simple measure of liquidity) is much smaller on Hotspot FX than has been reported by \cite{Wyart:2008relation} for the LSE and Paris Stock Exchange, which operate as centralized LOBs. Specifically, \cite{Wyart:2008relation} reported ratios in the range $100$--$1000$ for the stocks that they studied, and they argued that this provides strong evidence that available liquidity (in the form of limit orders) is generally in short supply. On Hotspot FX, the same ratios (see Table \ref{tab:sfxallagg}) vary between roughly $2$ (for EUR/GBP) and roughly $10$ (for EUR/USD). One simple explanation for this result is that liquidity is much more plentiful on Hotspot FX than is the case in other markets. Although this explanation is somewhat plausible, it seems unrealistic that the corresponding results for the different markets should differ by a factor of 50 or more. In a QCLOB, the appropriate quantity to assess the liquidity available to a given institution $\theta_i$ is not the mean total size of all active orders in the global LOB $\mathcal{L}(t)$, but rather the mean total size of active orders in $\theta_i$'s local LOB $\mathcal{L}_i(t)$. If the fraction of liquidity from $\mathcal{L}(t)$ available in $\mathcal{L}_i(t)$ is also small, then the corresponding ratio of the mean total size of market orders on a single day to the mean total size of active orders available to $\theta_i$ could be similar to the range reported by \cite{Wyart:2008relation} for centralized LOBs.

Our results suggest that institutions monitor $\mathcal{L}_i(t)$ carefully when deciding how to act. For example, we observe few market orders that match at several different prices (see Table \ref{tab:sfxallagg}). This suggests that many institutions implement selective liquidity-taking strategies by monitoring $\mathcal{L}_i(t)$ and only submitting market orders with a size smaller than the depth at $b_i(t)$ or $a_i(t)$. Correspondingly, we find that the mean size of market orders is less than half of the mean size of limit orders (see Table \ref{tab:sfxallagg}).

Our results suggest that trade-relative coordinates provide a useful perspective for studying QCLOBs. Naturally, there are some weaknesses with this approach: For example, an institution $\theta_i$ may not regard the most recent trade prices as particularly important if they deviate significantly from its local quotes $b_i(t)$ and $a_i(t)$. Moreover, the mean inter-arrival time for EUR/GBP market orders is more than 1 minute (see Table \ref{tab:sfxallagg}), so the values of $B(t)$ and $A(t)$ update relatively infrequently, yet our results suggest that some institutions act extremely quickly to capitalize on possible arbitrage opportunities that arise in their local LOB $\mathcal{L}_i(t)$. Together, these results suggests that institutions may regard the information in their local LOB to be more important when making quick-fire trading decisions on short timescales of seconds or milliseconds, but may regard the values of $B(t)$ and $A(t)$ to be more important when making less rapid trading decisions on longer time scales.

The slow updating of $B(t)$ and $A(t)$ may also be regarded as a benefit of trade-relative coordinates, because it ensures that price measurements are stable over time. The rise in popularity of electronic trading has led to a sharp increase in the frequency of order arrivals near the best quotes \cite{Biais:2011equilibrium,Chaboud:2011rise,Cont:2011statistical,Hendershott:2011algorithmic, Kirilenko:2011flash}, which cause the values of $b(t)$ and $a(t)$ --- and therefore the quote-relative prices of all orders --- to fluctuate rapidly. By contrast, trade-relative prices change only when a trade occurs, and they consequently avoid the difficulties caused by the extremely fast update frequency of the best quotes.

The strong round-number effects that we observe in the trade-relative distributions (see Figures \ref{fig:sfxLOdistday}, \ref{fig:sfxCXdistday}, \ref{fig:sfxDPdistday}, and \ref{fig:sfxCPOday}) suggest that institutions do indeed calculate and consider trade-relative prices. In centralized LOBs, quote-relative distributions often contain strong round-number periodicities at integer multiples of 10 ticks \cite{Challet:2001analyzing,Gu:2008empiricalshape,Mu:2009preferred,Zhao:2010model}. We find relatively weak evidence for this behaviour on Hotspot FX (see Figure \ref{fig:sfxFFT}). The strong periodicities that we observe in trade-relative coordinates are extremely unlikely to emerge by chance, so it seems that institutions regard $B(t)$ and $A(t)$ as important sources of information when deciding how to act.

In both quote-relative and trade-relative coordinates, the distributions of limit order arrivals, cancellations, and normalized mean depths on Hotspot FX exhibit considerable variation across different trading days (see Figure \ref{fig:sfxDailyECDFs}). In all cases, however, rescaling the data to account for differences in the first two moments significantly reduces the mean pairwise CvM distance between daily distributions (see Figure \ref{fig:sfxDailyECDFsRescaled} and Table \ref{tab:sfxoosperformance}). In trade-relative coordinates, the resulting curve collapse for limit order arrivals and cancellations is particularly strong. Given the turbulent macroeconomic activity that occurred during this period, such strong curve collapse is surprising, because it indicates that the first two moments provide significant explanatory power for daily order flow and highlights that the vast majority of daily variations in order flow appear to be linear transformations of a single, universal curve.

\section{Conclusions and Outlook}\label{sec:conclusions}

During the past decade, a rich and diverse literature has helped to illuminate many important aspects of trading via LOBs. To date, however, almost all work in this area has addressed only centralized LOBs, in which all institutions can trade with all others. In this paper, we have provided a detailed description of an alternative LOB configuration, which we call a QCLOB, and performed an empirical analysis of a recent, high-quality data set from a large electronic trading platform, Hotspot FX, which utilizes this mechanism to facilitate trade.

Our results reveal some important differences between QCLOBs and centralized LOBs. For example, we observed many instances in the Hotspot FX data where the global bid-ask spread was negative, whereas this is not possible in a centralized LOB. We also observed differences between the distributions of order flow and LOB state on Hotspot FX and the corresponding distributions reported by empirical studies of centralized LOB. These differences underline the need for detailed investigations of other widely used market organizations to complement the sizeable literature on centralized LOBs.

Our use of trade-relative coordinates illuminated several interesting properties of order flow and LOB state that are not apparent when measuring prices relative to the prevailing quotes, as is common when studying centralized LOBs. The strong round-number effects that we observed in trade-relative coordinates suggest that institutions trading on Hotspot FX regard the most recent trade prices as an important source of information when deciding how to act. Although our use of trade-relative coordinates was motivated by the structure of a QCLOB, we conjecture that this coordinate frame may also provide useful insight into centralized LOBs. It would be interesting to perform an empirical analysis of a centralized LOB in trade-relative coordinates to facilitate comparisons with our findings. To our knowledge, no such empirical studies have yet been conducted. We therefore believe this to be an interesting avenue for future research.

In a recent study of the LSE, \cite{Axioglou:2011markets} conjectured that the statistical properties of financial markets change every day. At present, however, many of the most widely discussed LOB models operate under the assumption that order flow is governed by stochastic processes with fixed rate parameters \cite{Challet:2001analyzing,Cont:2010stochastic,Farmer:2005predictive,Mike:2008empirical,Smith:2003statistical,Toth:2011anomalous}. The empirical verification of such models has typically consisted of comparing their output to long-run statistical averages from large data sets. Our results, together with those of \cite{Axioglou:2011markets}, bring into question the usefulness of using long-run statistical averages to forecast activity on a specific day. It would be interesting to study the performance of several existing LOB models to assess their performance on shorter timescales. Given that regulators require many institutions to make risk calculations on a daily basis, this is an important task for future research.

Finally, we note that our statistical analysis mainly examined aggregate order flow and the global LOB $\mathcal{L}(t)$. An interesting challenge for future research will be to gain a deeper understanding of the subset of liquidity in $\mathcal{L}(t)$ that individual institutions can access in their local LOBs. There are several aspects to this question -- including understanding the structure of the network of CCLs $c_{i,j}$ between individual institutions, understanding how $\mathcal{L}_i(t)$ varies across different institutions, and assessing how the restriction of trading opportunities to institutions with sufficient CCLs impacts price formation and market stability. We aim to address these, and many other related questions, in our forthcoming work.

\appendix

\section{Fitting the Generalized $t$ Distribution}\label{app:fittingt}

Let $Z$ be a random variable from the standard normal distribution, and let $V$ be an independent random variable from the chi-squared distribution with $\nu$ degrees of freedom. The random variable\begin{equation}\label{eq:gent}T = \sigma \frac{Z + \xi}{\sqrt{V/ \nu}} + \mu\end{equation}then follows a \emph{generalized $t$ distribution}. The parameters $\mu$, $\sigma$, and $\xi$ extend the classical Student's $t$ distribution by providing explicit control over the distribution's mean, variance, and skewness, respectively \cite{Gosset:1908probable}.

For each day $d \in \left\{1,2,\ldots,30\right\}$, we fit the generalized $t$ distribution to a given property of the Hotspot FX data (e.g., EUR/USD limit order arrivals in trade-relative coordinates) by minimizing the Cram\'{e}r--von Mises (CvM) distance \cite{Cramer:1928composition}\begin{equation}\label{eq:sfxcvmdist}C=\int_{p}\left[F_{d}(p)-F(p;\mu,\sigma,\xi,\nu)\right]^2 dF(p;\mu,\sigma,\xi,\nu)\end{equation}between the ECDF $F_{d}$ of the given property on day $d$ and the cumulative density function $F$ of the generalized $t$ distribution with parameters $\mu$, $\sigma$, $\xi$, and $\nu$. We use Newton's method \cite{Dennis:1983numerical} to optimize the objective function in Equation (\ref{eq:sfxcvmdist}) over these parameters. On a standard desktop computer with a 2GHz processor and 8GB RAM, this process requires approximately $1$--$2$ minutes of computation to fit the distribution of a given property for a given currency pair on a given day.

Fitting a distribution by minimizing the CvM distance is equivalent to minimizing a least-squares objective function that assigns more weight to the regions of the distribution with greater density. It is also possible to fit the generalized $t$ distribution via moment-matching \cite{Hall:2005generalized} or maximum-likelihood \cite{Casella:2001statistical} techniques, but the resulting estimates do not perform as well due to the existence of a handful of orders with extremely large relative prices that strongly impact the sample moments and maximum-likelihood estimates.

\section{Quantifying the Strength of Curve Collapse}\label{app:curvecollapse}

To quantify the strength of curve collapse from rescaling each day's data, we calculate the mean of the ratio of CvM distances (see Equation (\ref{eq:sfxcvmdist})) between the ECDFs of a chosen property on a given pair of trading days before and after applying the rescaling. More precisely, we calculate\begin{equation}\label{eq:Cbarsimple}\overline{C} = \frac{1}{30 \times 29}\sum_{\substack{d_1,d_2 \\ d_1 \neq d_2}} \frac{C^{(1)}_{d_1,d_2}}{C^{(2)}_{d_1,d_2}},\end{equation}where $C^{(1)}_{d_1,d_2}$ denotes the CvM distance between the ECDFs of a chosen property (e.g., EUR/USD limit order arrivals in quote-relative coordinates) on days $d_1$ and $d_2$, and $C^{(2)}_{d_1,d_2}$ denotes the CvM distance between the same ECDFs after rescaling the data on day $d_2$ by subtracting the mean for day $d_2$ and dividing by the standard deviation for day $d_2$, then multiplying the result by the standard deviation for day $d_1$, and finally adding the mean for day $d_1$. Larger values of $\overline{C}$ correspond to stronger collapse of the ECDFs. Note that we do not rescale the data from both days to measure the distance between the rescaled distributions directly, but we instead apply the inverse rescaling from day $d_1$ to the rescaled data from day $d_2$. This ensures that we measure our results in units of price for both $C^{(1)}_{d_1,d_2}$ and $C^{(2)}_{d_1,d_2}$, rather than using units of rescaled price for $C^{(2)}_{d_1,d_2}$.

\section*{Acknowledgements}We thank Bruno Biais, Jean-Philippe Bouchaud, Rama Cont, J. Doyne Farmer, Austin Gerig, Ben Hambly, Nikolaus Hautsch, Gabriele La Spada, Sergei Maslov, Stephen Roberts, Eric Schaanning, Torsten Sch\"{o}neborn, Cosma Shalizi, Thaleia Zariphopoulou, and Ilija Zovko for useful discussions. We thank Hotspot FX for providing the data for this project, and we thank Jonas Haase, Terry Lyons, Rich Plummer-Powell, and Justin Sharp for technical support. We also thank two anonymous reviewers for many helpful comments and suggestions. MDG and SDH thank the Oxford-Man Institute of Quantitative Finance, and MDG thanks EPSRC and the James S. McDonnell Foundation for supporting this research.

\bibliographystyle{plain}

\bibliography{../../dphilLOBbib}

\end{document}